\documentclass[12pt]{article}
\usepackage{epsf}
\DeclareMathAlphabet{\mib}{OML}{cmm} {b}{it}
\makeatletter

\@addtoreset{equation}{section}
\makeatother
\def\gtsim{\mathrel{\hbox{\raise0.2ex
\hbox{$>$}\kern-0.75em\raise-0.9ex\hbox{$\sim$}}}}
\def\ltsim{\mathrel{\hbox{\raise0.2ex
\hbox{$<$}\kern-0.75em\raise-0.9ex\hbox{$\sim$}}}}
\def\llt{\mathrel{<\kern-0.5em <}}   %  doubled inequality  <<
\def\ggt{\mathrel{>\kern-0.5em >}}   %  doubled inequality  >>

\def\ket#1{| #1 \rangle}

\def\expecv#1{\langle #1 \rangle}
\def\absv#1{\left|#1\right|}
\def\Tr{{\rm Tr}}
\def\del{{\partial}}
\def\Im{{\rm Im}}       % imaginary part
\def\half{{1\over2}}
\def\kslash{k\kern-0.55em\raise 0.14ex\hbox{/}}
\def\Aslash{A\kern-0.6em\raise 0.14ex\hbox{/}}
\def\dslash{\del\kern-0.55em\raise 0.14ex\hbox{/}}
\def\sinb{{\sin\beta}}
\def\cosb{{\cos\beta}}
\newcommand{\PRD}[3]{Phys.~Rev. {\bf D{#1}} (19{#2}) {#3}}
\newcommand{\PRL}[3]{Phys.~Rev.~Lett. {\bf {#1}} (19{#2}) {#3}}
\newcommand{\NPB}[3]{Nucl.~Phys. {\bf B{#1}} (19{#2}) {#3}}
\newcommand{\PLB}[3]{Phys.~Lett. {\bf B{#1}} (19{#2}) {#3}}
\newcommand{\PTP}[3]{Prog.~Theor.~Phys. {\bf {#1}} (19{#2}) {#3}}
\newcommand{\ANN}[3]{Ann.~Phys. {\bf {#1}} (19{#2}) {#3}}
\begin{document}
\begin{titlepage}
\begin{flushright}
SAGA--HE--108\\
August 16,~1996
\end{flushright}
\vspace{24pt}
\begin{center}{\Large\bf
$CP$ Violation and Baryogenesis\\
at the Electroweak Phase Transition
}
\end{center}
\vspace{24pt}
\begin{center}{\bf
Koichi Funakubo\footnote{e-mail: funakubo@cc.saga-u.ac.jp} 
}\end{center}
\vskip 1.0cm
\begin{center}{\sl
Department of Physics, Saga University, Saga 840
}\end{center}
\vskip 1.0cm
\centerline{\bf Abstract}
\vskip 0.2 cm
We report a recent attempt to relate $CP$ violation in the electroweak
theory with two Higgs doublets to the baryon asymmetry generated at
the electroweak phase transition, after surveying the scenario of
the electroweak baryogenesis.
\end{titlepage}
\baselineskip=18pt
\setcounter{page}{2}
\setcounter{footnote}{0}
%
%
%-------------------
\section{Introduction}
The evolution of the universe is well described by the standard hot
big bang model and its early stage offers a common field for
particle physics and astrophysics\cite{rev-cosmology}. 
Not only have various observations in laboratories been clues to
understand what occurred in the early universe,
but new ideas have also led to resolutions of the problems
concerning very early universe; {\rm e.g.}, grand unified models 
with strong first-order phase transitions led to the idea of inflation,
and dark matter
may be explained by new physics, which will be examined in the near future.
On the other hand, astrophysical observations give constraints on 
models of particle physics which cannot be obtained from accelerator 
experiments\cite{PDG}.\ 
The baryon asymmetry of the universe (BAU) is among these topics.\par
The BAU is one of the most obvious facts\cite{rev-cosmology}:\   
we do not observe any antimatter in our solar system, and 
high energy cosmic rays, which contain a small amount of anti-protons
consistent as secondary products, are evidence of the baryon asymmetry
on the galactic scale. The absence of hard $\gamma$ rays, which would be 
emitted on nucleon-antinucleon annihilation, from nearby 
clusters of galaxies, such as the Virgo cluster, implies that such a 
cluster consists of $(1\sim100)M_{\rm galaxy}\simeq10^{12\sim14}M_\odot$
of either baryons or antibaryons only.
While we have no evidence for baryon asymmetry on a larger scale,
matter should be separated from antimatter on scale of, at least,
$10^{12}M_\odot$.
We characterize this manifest quantity by the ratio of the baryon 
number to entropy
\begin{equation}
 {{n_B}\over s} \equiv {{n_b - n_{\bar b}}\over s}, \label{eq:def-BAU}
\end{equation}
where $s$ is the entropy density and $n_b$($n_{\bar b}$) is the 
(anti)baryon number density. This remains constant in the absence of a
baryon-number-changing process and entropy production, during the 
expansion of the universe. 
To explain the light-element abundances within the framework of the 
standard big-bang nucleosynthesis, it is required that
$\eta\equiv n_B/n_\gamma=(1.5-6.3)\times10^{-10}$\cite{PDG}.\  
As shown in Appendix~A, the photon density $n_\gamma$ is related to $s$
by $s=7.04n_\gamma$ at present\cite{rev-cosmology},\  so that
\begin{equation}
 {{n_B}\over s} = (0.21-0.90)\times10^{-10}.  \label{eq:value-BAU}
\end{equation}
This is very small but sufficient to form the present visible matter.
Although one might think that matter and antimatter were separated on
a scale of $10^{12}M_\odot$ in locally baryon-symmetric universe,
the present BAU cannot be explained without any primordial asymmetry.
The present baryons are those which survived 
the nucleon-antinucleon annihilation and which froze out at
$T\simeq 20\mbox{MeV}$. Since starting from the baryon-symmetric universe 
$n_b/s\simeq n_{\bar b}/s\sim 7\times10^{-20}$ after the freeze out,
the BAU, which is about nine orders larger, could not be 
generated\cite{rev-cosmology},\  
To avoid this annihilation, suppose that some hypothetical mechanism
separated nucleons and antinucleons before $T\simeq 38\mbox{MeV}$,
when $n_b/s= n_{\bar b}/s\sim 8\times10^{-11}$.
At that time the matter contained in the causal region was only about
$10^{-7}M_\odot\llt 10^{12}M_\odot$.
Hence it is natural to assume that the universe had baryon asymmetry
before it cooled down to $T\simeq 38\mbox{MeV}$.\par
To obtain the BAU starting from the symmetric universe,
three conditions, first proposed by Sakharov\cite{Sakharov},\  
must be satisfied:
\begin{itemize}
 \item[(1)] baryon number violation,
 \item[(2)] $C$ and $CP$ violation,
 \item[(3)] departure from equilibrium.
\end{itemize}
Without condition (1), the symmetric universe remains 
baryon-symmetric, and it seems difficult to realize a
locally asymmetric but globally symmetric universe, as we saw above.
If $C$ and $CP$ were conserved, baryon number could not be generated in the 
symmetric universe. This is understood as follows. Suppose that
the initial state of the universe is described by a density operator
$\rho_0$ which is $C$- and $CP$-invariant.
Since baryon number is odd under $C$ and $CP$, 
$\expecv{n_B}_0=\Tr[\rho_0 n_B]=0$. If the hamiltonian of the system is
$C$- {\it or} $CP$-invariant, the density operator $\rho$ at later time,
whose time evolution is governed by the Liouville equation, is also
invariant under $C$ {\it or} $CP$ transformation: $[\rho,{\cal C}]=0$ {\it or}
$[\rho, {\cal CP}]=0$. Because of ${\cal C}B{\cal C}^{-1}=-B$ and
$[B,{\cal P}]=0$,
$\expecv{n_B}=\Tr[\rho n_B]=\Tr[\rho\,{\cal C}n_B{\cal C}^{-1}]=-\Tr[\rho n_B]$,
{\it or} $\expecv{n_B}=\Tr[\rho\,{\cal CP}n_B({\cal CP})^{-1}]=-\Tr[\rho n_B]$.
Here ${\cal C}$(${\cal P}$) is the operator representing the charge 
conjugation (space inversion).
In any case we have $\expecv{n_B}=0$. Hence {\it both} $C$ {\it and} 
$CP$ must be violated to have nonzero $\expecv{n_B}$.
If baryon-number-changing processes are in chemical equilibrium,
the chemical potential for the baryon number vanishes so that
the equilibrium distribution of baryons coincides with that of
antibaryons: $n_b = n_{\bar b}$.
This implies that even if the universe was baryon-asymmetric at some 
time, the BAU vanishes after the universe experienced an equilibrium 
era when baryon-number changing processes were in effect.\par
As the models of particle interactions satisfying conditions (1) 
and (2), we now have electroweak theories, grand unified theories 
(GUTs), supersymmetric extensions of them and others.
Among these, baryogenesis within GUTs was first extensively 
studied\cite{GUT-BAU,rev-cosmology}.\  
In the framework of GUT-baryogenesis, the main process with baryon 
number violation is
the out-of-equilibrium decay of the heavy bosons $X$, which are the gauge
or Higgs bosons of mass $m_X\gtsim10^{15}\mbox{GeV}$.
Suppose that $X$ decays into two channels $qq$ ($\Delta B=2/3$) and 
$\bar q\bar l$ ($\Delta B=-1/3$) with branching ratio $r$ and $1-r$,
respectively.
$C$ and $CP$ are violated if $r$ is not equal to the branching
ratio $\bar r$ of the process ${\bar X}\rightarrow{\bar q}{\bar q}$. 
Then the expectation value of the change in baryon number in the decay
of $X$-$\bar X$ pairs is
$\expecv{\Delta B}={2\over3}r-{1\over3}(1-r) - {2\over3}\bar r
 +{1\over3}(1-\bar r)=r-\bar r$.
If $C$ {\it or} $CP$ is conserved, $r=\bar r$ so that $B$ is not
generated.\footnote{Here we do not see spin and momentum in the final
states, so that $P$ does not affect $r$.}\  At $T\simeq m_X$, the decay
rate of $X$ bosons is roughly given by $\Gamma_D\simeq\alpha m_X$,
where $\alpha=g^2/(4\pi)$ with $g$ being the coupling constant.
($\alpha\sim1/40$ for the gauge boson, $\alpha\sim10^{-6\sim-3}$
for the Higgs boson.)
The Hubble parameter is $H\sim 1.7\sqrt{g_*}T^2/m_{Pl}$, where
$g_*\simeq10^{2\sim3}$ is the effective massless degrees of freedom.
(See Appendix A.) At temperatures near $m_X$, $\Gamma_D\simeq H$,
so that the chemical equilibrium between the decay and the production 
of $X$ and $\bar X$ no longer holds.
The Boltzmann equations are used to estimate the generated baryon
number quantitatively\cite{GUT-BAU-review}.\par
Since the weak interaction incorporates $C$ and $CP$ violation, it and its
extensions may be candidates to offer the BAU if they fulfill
the other conditions (1) and (3).
It is well known that in the standard model $B+L$ is violated by
the axial $U(1)$ anomaly.
This anomalous process can convert primordial $L$ into $B$.
Hence if $L$ is generated by some $L$-violating interaction at the
intermediate scale between the electroweak and GUTs scales, 
$B$ would be left after the electroweak phase transition.
This might be another candidate for the baryogenesis.
If the $B$-violating process occurs out of equilibrium at the
electroweak scale,
the BAU might be generated within the framework of the electroweak
theory. This possibility is rather attractive, since it depends only 
on physics which could be tested by near future experiments. 
In this article, we review the attempt to realize this idea ---
electroweak baryogenesis, focusing on the relation between the BAU and
$CP$ violation in the Higgs sector of the extensions of the standard model.
For earlier works on this subject, see the 
review article, Ref.~\cite{CKN-review}.\    
This paper is organized as follows.
In \S~2, we present the basic idea of the electroweak baryogenesis.
In the following three sections, we study how it is realized in some
detail. We summarize recent attempts to determine $CP$ violation at
the electroweak phase transition in \S~6.
The final section is devoted to concluding remarks.
%
%
%----------
\section{Overview of the electroweak baryogenesis}
The axial anomaly in the electroweak theory nonperturbatively violates
the sum of baryon number and lepton number
$B+L$ and its probability is enhanced at high temperatures in the symmetric
phase of the gauge symmetry, while suppressed at low temperatures in
the broken phase, as we shall see in the next section.
Near the electroweak phase transition (EWPT) temperature
$\sim100\mbox{GeV}$, the universe was expanding
too slowly to make the anomalous process out of equilibrium. 
If the EWPT is first order accompanying formation and growth of bubbles
of the broken phase in the symmetric phase, the $B+L$-violating process 
will deviate from equilibrium near the bubble walls.
$C$ and $CP$ violation of electroweak theory may distinguish the effects
of the expanding bubble on the fermions and antifermions leading to
the generation of $B+L$. If the anomalous process is fully suppressed in
the broken phase, the BAU is left after the EWPT.
We shall refer to this scenario as `electroweak baryogenesis'.
To investigate the possibility of this idea, we need a wide range of
knowledge in theoretical physics;
estimation of the nonperturbative $B+L$ violating ratio,
finite-temperature field theory to determine the static properties
of the EWPT, dynamics of the phase transition and model building
of electroweak theory consistent with present experiments.\par
Before surveying each subject, we briefly introduce other attempts
to produce the BAU by use of the anomalous $B+L$ violation.
Since $B+L$ rather than $B$ is violated, $B+L$ is washed out if the 
anomalous process was in equilibrium. This implies that we need primordial
$B-L$ to have the present BAU if the anomalous process had been in 
equilibrium until it froze out\cite{KRS}.\  (We shall discuss this
issue in the next section.)
Then the GUTs which conserve $B-L$, such as the $SU(5)$ model, would be 
useless.
As the origin of nonzero primordial $B-L$, one may consider a GUT which
does not conserve this quantity. 
This assumes new physics at the GUT scale 
($\sim10^{16}\mbox{GeV}$), which could be tested by the proton decay 
experiments.
The $L$-violation at the intermediate scale between the electroweak and
GUTs scales with the Majorana neutrino might seed the primordial
$B-L$\cite{Fukugita}.\   For the $L$-violating process not to erase $L$
completely, some conditions on the masses are imposed, which could
be checked by solar neutrino and other experiments.
Supersymmetric extensions of the standard model contain the
scalar superpartners with the same internal quantum number as
the quarks and leptons.
These scalar fields could have nonzero expectation values along the
`flat directions', which are typical in such models, 
leading to $B$ and/or $L$ violation at high temperatures.
This mechanism is investigated in Refs.\cite{AD} and would be
a probable origin if supersymmetry is discovered in the experiments.
In addition to these, topological defects, such as strings and domain walls,
formed at the electroweak scale, could create the baryon number when
they move or decay\cite{EWstring}.\par
The present BAU might be generated by one or some combination of these
mechanisms, including the electroweak baryogenesis.
In any case, the electroweak baryogenesis would be the last chance to
generate the BAU if the anomalous process froze out after the EWPT,
so that it might affect the baryon number already created until that time.
Once a model of the electroweak theory is specified, the nature of 
$CP$ violation and the EWPT are, in principle, known, then the BAU 
generated can be evaluated. When quantitative study of the electroweak
baryogenesis is developed, one can select a model to explain the 
present BAU and such a model can be confirmed by experiments
in the near future.
This is why much attention is paid to this subject and related
subjects such as the finite-temperature phase transition of the 
gauge-Higgs system and the dynamics of the first-order phase 
transition.
%
%
%
%-----------
\section{Sphaleron process}
\subsection{Anomalous fermion number nonconservation}
In the standard model and its extensions with the same fermion-gauge
interaction, the current of $B+L$, suffers from the axial
anomaly\cite{anomaly-rev}:
\begin{eqnarray}
 \del_\mu j^\mu_{B+L}&=&{N_f\over{16\pi^2}}[g^2\Tr(F_{\mu\nu}\tilde F^{\mu\nu})
                        - {g^\prime}^2 B_{\mu\nu}\tilde B^{\mu\nu} ],
                                   \label{eq:anomaly1}  \\
 \del_\mu j^\mu_{B-L} &=& 0,       \label{eq:anomaly2}
\end{eqnarray}
where $N_f$ is the number of the generations, $g$ ($g^\prime$) and
$F_{\mu\nu}$ ($B_{\mu\nu}$) are the gauge coupling and the field strength 
of the $SU(2)_L$ ($U(1)_Y$) gauge field $A_\mu(x)$
($B_\mu(x)$), respectively, and the tilde denotes the Hodge dual,
$\tilde F^{\mu\nu}\equiv{1\over2}\epsilon^{\mu\nu\rho\sigma}F_{\rho\sigma}$.
Integrating these equations over
$(t,{\mib x})\in[t_i,t_f]\otimes{\bf R}^3$, we obtain
\begin{eqnarray}
 B(t_f)-B(t_i) &=&
 {{N_f}\over{32\pi^2}}\int_{t_i}^{t_f}d^4x\,\left[
         g^2\Tr(F_{\mu\nu}\tilde F^{\mu\nu})
      - {g^\prime}^2 B_{\mu\nu}\tilde B^{\mu\nu}\right]  \nonumber\\
 &=& N_f\left[N_{CS}(t_f) - N_{CS}(t_i)\right],     \label{eq:CSdiff}
\end{eqnarray}
where $N_{CS}$ is the Chern-Simons number defined, in the Weyl gauge
($A_0=0$), by
\begin{equation}
  N_{CS}(t) = {1\over{32\pi^2}}\int d^3x\,\epsilon_{ijk}\left[
              g^2\Tr\left(F_{ij}A_k-{2\over3}gA_iA_jA_k\right)
              -{g^\prime}^2 B_{ij}B_k \right]_t.
                                   \label{eq:def-CS}
\end{equation}
The Chern-Simons number is not gauge-invariant but its difference
(\ref{eq:CSdiff}) is apparently gauge-invariant.
For a pure-gauge configuration, $F_{ij}=B_{ij}=0$, $N_{CS}$ takes
an integer value. That is, the classical vacua of the gauge sector
are labeled by the integer $N_{CS}$, which is the winding number
corresponding to the homotopy group $\pi_3(SU(2))\simeq{\bf Z}$,
since the $U(1)$ contribution in (\ref{eq:def-CS}) always vanishes
for a pure gauge. In the semiclassical approximation, 
the vacuum-to-vacuum amplitude with $\Delta N_{CS}=1$ in the
$SU(2)$ gauge system is dominated by the instanton
configuration\cite{BPST},\  which is a classical solution with
finite euclidean action $S_{\rm inst}=8\pi^2/g^2$, and is given
by $\sim\exp(-S_{\rm inst})$\cite{tHooft}.\  
For a strong-coupling theory such as QCD, this amplitude is so large
that the quantum vacuum is a superposition of these degenerate 
classical vacua known as the $\theta$-vacuum\cite{JackiwRebbi}.\  
As for the $SU(2)$ gauge-Higgs system, there is no classical solution with
a finite euclidean action, but the constrained 
instanton\cite{constrained}\  
or the valley instanton\cite{valley}\  plays a similar role and
yields the same suppression $\sim\exp(-S_{\rm inst})$ together
with other factors.
For the standard model, this suppression factor is so small,
$\exp(-S_{\rm inst})=\exp(-2\pi\sin^2\theta_W/\alpha_{em})
\simeq{ e}^{-189}=10^{-82}$, that the
vacuum-to-vacuum transition with $\Delta N_{CS}=1$ is not observable.
That is, $B+L$ can hardly change in the vacuum at zero temperature.\par
The nonconservation of the current (\ref{eq:anomaly1}) indicates
that $B+L$ associated with the quarks and leptons actually
changes when the background gauge fields have nontrivial topology.
The Atiyah-Singer index theorem says that the integral
$ {g^2\over{32\pi^2}}\int d^4 \Tr[F\tilde F]$ --- the Pontrjagin index ---
is related to the zero mode of the chiral fermion\cite{Fujikawa}.\ 
This can be seen more directly if one considers the Dirac equation
in this background of the gauge fields. 
Regarding the euclidean time as a parameter, the solution to
the Dirac equation for the left-handed fermion in the instanton
background shows that the level in the Dirac sea moves to a 
positive-energy level, as the parameter goes from $-\infty$ to
$\infty$\cite{levelx}.\  This is known as the level crossing 
phenomenon.\par
\subsection{Sphaleron transition}
\begin{figure}
 \epsfxsize = 7cm
 \centerline{\epsfbox{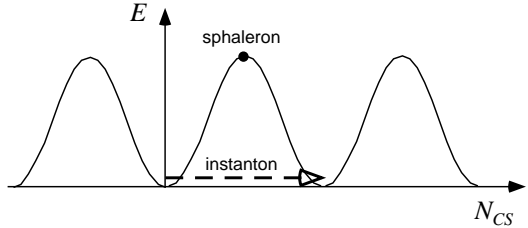}}
 \caption{Schematical view of the energy as a functional of
 the field configuration. The horizontal axis denotes the 
 Chern-Simons number, which is a subspace of the configuration space.}
 \label{fig:1}
\end{figure}

Our main concern is whether the anomalous process is enhanced at
high temperatures. 
This possibility was first suggested when the energy barrier between
the neighboring classical vacua labeled by $N_{CS}$ was found to
be finite.
The top of the barrier in Fig.~\ref{fig:1} corresponds to the saddle point
in the configuration space, which is a static saddle-point solution
with finite energy of the $SU(2)$ gauge-Higgs system\cite{sph-sol1}.\  
Such a solution is called a ``sphaleron''.
%
%%%%%%% Fig.1 %%%%%%%%
In contrast to a topological soliton such as nonabelian monopoles
and Skyrmions, it is unstable since the fluctuation spectrum around
the solution contains one negative mode. It is a nontopological 
solution and has $N_{CS}=1/2$ in some gauge, in which the vacua
divided by it have $N_{CS}=0$ and $1$. The negative mode represents
the instability against decay to the neighboring vacua.
Although a `sphaleron' is not a topological soliton, its existence
is related to the topology of configuration space of a field theory;
some class of field theories having `noncontractible loop' or
its higher dimensional generalization may have such a saddle-point
solution\cite{sph-sol1,sph-sol3}.\  
Among these theories are the $1+1$-dimensional $U(1)$ gauge-Higgs
system\cite{sph-sol4}\  and the $1+1$-dimensional $O(3)$ nonlinear sigma
model\cite{sph-sol5},\  both of which also have instanton solutions.
The static energy of the sphaleron in the system of $SU(2)$ gauge
fields and a Higgs doublet is given by
\begin{equation}
  E_{\rm sph}={{2M_W}\over{\alpha_W}}B\left({\lambda\over{g^2}}\right)
         \simeq 10\mbox{TeV},      \label{eq:sph-energy}
\end{equation}
where $\lambda$ is the Higgs self coupling, $\alpha_W=g^2/(4\pi)$ and
$1.5\le B\le 2.7$ for $\lambda/g^2\in[0,\infty)$\cite{sph-sol2}.\  
The sphaleron solution in the standard model with nonzero $\theta_W$
has been found and its energy is somewhat smaller\cite{sph-SM}.\par
For temperatures below the barrier height, the transition rate
between the classical vacua was estimated semiclassically by 
Affleck\cite{Affleck},\  whose classical statistical version had been
formulated by Langer\cite{Langer}.\  
Affleck showed, in the WKB approximation of a quantum mechanical problem
with metastable potential, that for lower temperature than
$\omega_-/(2\pi)$, where $\omega_-$ is the negative-mode frequency at
the top of the barrier, the transition rate is given by
$\Gamma\simeq 2\,\Im\,F$,
while for $T\gtsim \omega_-/(2\pi)$,
\begin{equation}
 \Gamma \simeq {{\omega_-}\over{\pi T}}\Im\,F.   \label{eq:Gamma}
\end{equation}
Here the free energy $F\equiv-T\log\Tr[{ e}^{-H/T}]$, expressed
in the path-integral form, should be estimated around the dominant
configurations, the bounce\cite{CallanColeman}\  
at low temperatures and the top of the barrier (sphaleron) at
high temperatures. In any case, the `imaginary part' arises from
the Gaussian integral around each configuration, which contains 
a contribution from one negative mode.
At $T=0$, in the limit that the metastable vacuum becomes degenerate
with the stable one, the bounce action approaches twice
the instanton action, so that $\Gamma\sim\exp(-2S_{\rm inst})$,
reproducing the instanton calculation.
In field theories, the Gaussian integral contains contributions
from positive modes as well as zero modes corresponding to
global symmetries, such as translation and isospin, which are violated
by the configuration.\par
The first estimation of the transition rate in the $SU(2)$ gauge-Higgs
system was made by Arnold and McLerran\cite{Arnold}\  and is given by
\begin{equation}
 \Gamma_{\rm sph}^{(b)}
 \simeq k{\cal N}_{\rm tr}{\cal N}_{\rm rot}{{\omega_-}\over{2\pi}}
         \left({{\alpha_W(T) T}\over{4\pi}}\right)^3
         { e}^{-E_{\rm sph}/T},     \label{eq:sph-rate-br}
\end{equation}
where $\alpha_W(T)$ is the temperature-dependent fine structure constant, 
${\cal N}_{\rm tr}$ (${\cal N}_{\rm rot}$) is the zero-mode contribution
from translation (rotation), which is about $26$ ($5.3\times10^3$) in the
case $\lambda=g^2$, and $k$ denotes the contributions from the other 
modes. Here $\omega_-^2\simeq (1.8\sim6.6)m_W^2$ for
$10^{-2}\le\lambda/g^2\le10$ and $k$ is shown to be 
$O(1)$\cite{freeEsph}.\par
All of these results are valid in the existence of the mass scale
set by the vacuum expectation value of the Higgs, $v$.
At $T>T_C$, where $T_C$ is the EWPT temperature, $v(T)=0$,
so that the above results can no longer be applied.
For such high temperatures, we have no theory but on the grounds of
dimensional analysis\cite{Arnold},\  we have
\begin{equation}
 \Gamma_{\rm sph}^{(s)}\simeq \kappa(\alpha_W T)^4.  \label{eq:sph-rate-sym}
\end{equation}
Although there is no sphaleron solution in the symmetric phase,
we use the terminology `sphaleron transition' for the anomalous
process.
The formula (\ref{eq:sph-rate-sym}) was checked by classical Monte Carlo
simulations.
The outline of the simulation is as follows: First, an
ensemble of the coordinate and momentum of the classical field on a 
cubic lattice is generated by usual Metropolis or heat bath methods,
according to the statistical weight $\exp(-H/T)$, where $H$ is the classical 
hamiltonian. To avoid the Rayleigh-Jeans instability peculiar to
thermodynamics of classical fields, the finite lattice spacing
is adjusted at each temperature.
Picking up one of the configurations in this ensemble as the initial
condition, the classical equations of motion are solved numerically.
Thus $\expecv{N_{CS}^2(t)}$ is measured and is fitted to the expression
for the random walk, $\expecv{N_{CS}^2(t)}=2\Gamma Vt$ as 
$t\rightarrow\infty$. This program was executed for a $3+1$-dimensional
$SU(2)$ gauge-Higgs system in the symmetric phase  
and produced the result $\kappa>0.4$\cite{Ambjorn1},\
and for an $SU(2)$ pure gauge system\cite{Ambjorn2},\  
which may be good approximation of the symmetric phase of the 
gauge-Higgs system, and produced the result
$\kappa=1.09\pm0.04$.\footnote{Whether
the system is in the symmetric or broken phase can be checked by 
monitoring the expectation value of a gauge-invariant operator such
as $\Phi^\dagger(n)\Phi(n)$ and $\Phi^\dagger(n)U_\mu(n)\Phi(n+\mu)$.}
\subsection{Washout of $B+L$}
As we noted in the previous section, an unsuppressed sphaleron 
transition is needed for the electroweak baryogenesis, but
any generated baryon number may be washed out if the EWPT is
second order or the sphaleron transition does not decouple
after the EWPT\cite{KRS}.\  
For this decoupling to occur, the sphaleron rate should be smaller
than the expansion rate of the universe.
As we shall see in the next section, the EWPT took place at 
$T=T_C\simeq100\mbox{GeV}$. At this temperature, the Hubble parameter
is given by (see Appendix A)
\begin{equation}
 H(T_C)\simeq 1.7\sqrt{g_*}{{T_C^2}\over{m_{Pl}}}\simeq 
 10^{-13}\mbox{GeV},                           \label{eq:HatTc}
\end{equation}
where $g_*\sim100$ is the effective massless degrees of freedom at
this temperature. At $T>T_C$, since the sphaleron transition rate
per unit time, which is $\Gamma_{\rm sph}^{(s)}$ in (\ref{eq:sph-rate-sym})
multiplied by the particle density at that time, is much larger than
the expansion rate,
\begin{equation}
 \Gamma_{\rm sph}\simeq \kappa\alpha_W^4 T\sim10^{-4}\ggt H(T_C),
                            \label{eq:compare-rate}
\end{equation}
the $B+L$-changing process is in equilibrium in the symmetric phase.
As we show later, if the Higgs mass is larger than some value,
even the sphaleron rate in the broken phase is larger than the Hubble
parameter, so that the primordial $B+L$ is washed out.\par
The relic baryon number after this washout is estimated by use
of the relations among chemical potentials for the particles in the 
electroweak theory\cite{Harvey}.\  
The particle number density for each degree of freedom
at equilibrium is related to its chemical potential by
\begin{equation}
 n_+-n_-=\int{{d^3{\mib k}}\over{(2\pi)^2}}\left[
         {1\over{{ e}^{\beta(\omega_k-\mu)}\mp1}}
         - {1\over{{ e}^{\beta(\omega_k+\mu)}\mp1}}\right]
   \simeq\left\{\begin{array}{ll}
                \displaystyle{
                {{T^3}\over3}\,{\mu\over T}}\qquad\mbox{for bosons}\\
                \displaystyle{
                {{T^3}\over6}\,{\mu\over T}}\qquad\mbox{for fermions},
                \end{array}   \right.
          \label{eq:n-mu}
\end{equation}
where $\omega_k=\sqrt{{\mib k}^2+m^2}$ and we used $m/T\llt1$ and
$\mu/T\llt1$.
Since the elementary processes in the standard model are in chemical
equilibrium at high temperatures, there are relations among the chemical
potentials. For example, the equilibrium of the gauge interaction
imposes $\mu_W=\mu_{d_L}-\mu_{u_L}=\mu_{iL}-\mu_i=\mu_-+\mu_0$,
where $\mu_W$ is the chemical potential for $W^-$, 
$\mu_{u_L}$ ($\mu_{d_L}$) for the left-handed up-type (down-type) quarks,
$\mu_{iL,R}$ ($\mu_i$) for the $i$-th generation charged lepton (neutrino),
and $\mu_-$ ($\mu_0$)
for the charged Higgs $\phi^-$ (the neutral Higgs $\phi^0$).\footnote{
Since the strong interaction is in equilibrium, the chemical potential
for each flavor is independent of color. Further, the mixing among
the quark flavors is assumed to be in equilibrium.}
Consider the electroweak theory with $N_f$ generations and $m$ Higgs
doublets. When the sphaleron process such as
$\ket0\leftrightarrow u_Ld_Ld_L\nu_L$ is in equilibrium,
$ N_f(\mu_{u_L}+2\mu_{d_L})+\sum_i\mu_i = 0$.
Various quantum number densities, such as $B$, $L$, the electric 
charge $Q$, and the isospin ${\mib I}$ are expressed in terms of these
chemical potentials of the particles. By use of the relations among
the particle chemical potentials, one can express the quantum number
densities in terms of $\mu_W$, $\mu_{u_L}$, $\mu_0$ and 
$\mu=\sum_i\mu_i$:
\begin{eqnarray}
 B &=& N_f(\mu_{u_L}+\mu_{u_R}+\mu_{d_L}+\mu_{d_R})
    =  4N_f\mu_{u_L} + 2N_f\mu_W,       \label{eq:eq-B-mu}\\
 L &=& \sum_i(\mu_i+\mu_{iL}+\mu_{iR})
    =  3\mu + 2N_f\mu_W - N_f\mu_0,     \label{eq:eq-L-mu}\\
 Q &=& {2\over3}N_f(\mu_{u_L}+\mu_{u_R})\cdot3
      -{1\over3}N_f(\mu_{d_L}+\mu_{d_R})\cdot3      \nonumber\\
   & &\qquad -\sum_i(\mu_{iL}+\mu_{iR}-2\cdot2\mu_W-2m\mu_-\nonumber\\
   &=& 2N_f\mu_{u_L}-2\mu-(4N_f+4+2m)\mu_W+(4N_f+2m)\mu_0,
                                     \label{eq:eq-Q-mu}\\
 I_3&=& {1\over2}N_f(\mu_{u_L}-\mu_{d_L})\cdot3
       +{1\over2}\sum_i(\mu_i-\mu_{iL}) - 2\cdot2\mu_W
       -2\cdot{1\over2}m(\mu_0-\mu_-)            \nonumber\\
    &=& -(2N_f+m+4)\mu_W,              \label{eq:eq-T3-mu}
\end{eqnarray}
where we normalized the densities by $T^2/6$.
In the symmetric phase ($T>T_C$), all the gauge symmetries are 
respected so that $Q=I_3=0$, which constrain two of the remaining
chemical potentials. Then we have
\begin{equation}
 B = {{8N_f+4m}\over{22N_f+13m}}(B-L),\qquad
 L = -{{14N_f+9m}\over{22N_f+13m}}(B-L).   \label{eq:BL-sym}
\end{equation}
On the other hand, in the broken phase ($T<T_C$), because of
$Q=0$ and $\mu_0=0$, we have, if the sphaleron process is in
equilibrium,
\begin{equation}
 B = {{8N_f+4m+8}\over{24N_f+13m+26}}(B-L),\qquad
 L = -{{16N_f+9m+18}\over{24N_f+13m+26}}(B-L).   \label{eq:BL-br}
\end{equation}
These equations imply that if the primordial $B-L$ is absent,
we have no baryon number and lepton number.
Hence to have nonzero BAU after the EWPT, starting from the 
baryon-symmetric universe, 
\begin{itemize}
 \item[(i)] we must have $B-L$ before the sphaleron process 
 decouples, or
 \item[(ii)] $B+L$ must be created at the first-order EWPT, and
 the sphaleron process must decouple immediately after that.
\end{itemize}
Case (i) may be realized by mechanisms, stated in the previous 
section, such as GUTs, models with Majorana neutrinos, and the 
Affleck-Dine mechanism. The possibility (ii) is our main subject.\par
It should be noted that to derive the above relations, we neglected
the effects of mass in (\ref{eq:n-mu}).
In the standard model and its minimal supersymmetric extension,
this mass effect can leave nonzero BAU starting with vanishing
primordial $B-L$, if the model satisfies some
constraints on the lepton-generation mixing or $R$-parity 
violation\cite{DreinerRoss}.
%
%
%
%
%-------------
\section{Electroweak phase transition}
Gauge symmetries broken by expectation values of scalars in
nontrivial representations of the gauge group are known to be
restored at high temperatures\cite{Kirzhnits}.\  
The nature of the phase transition depends on the parameters in the
theory. Our aim is to clarify the properties of the EWPT, especially
its dynamics.
As is well known from the phase transition of classical systems
such as spin systems and liquid-vapor phase transition,
the dynamics of phase transitions are studied by use of the free
energy as a function of temperature and order parameters.
Fist we shall review analytic and numerical studies of the EWPT,
which focus on equilibrium properties, and then some attempts to
understand the dynamics of the EWPT. 
These are indispensable to determine whether the electroweak 
baryogenesis is possible and to estimate generated baryon number
quantitatively.
\subsection{Static properties of the phase transition}
\begin{figure}
 \epsfxsize = 8cm
 \centerline{\epsfbox{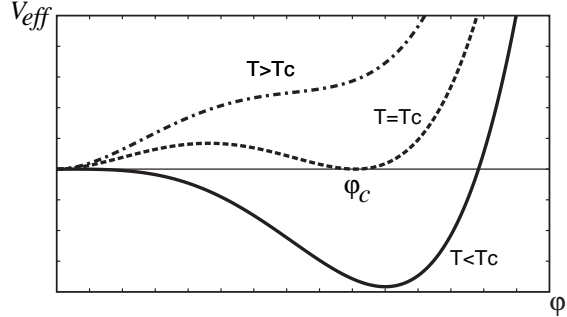}}
 \caption{Qualitative form of the effective potential at several 
 temperatures. The potential at $\varphi=0$ is normalized to be zero
 for all temperatures.}
 \label{fig:2}
\end{figure}

As noted in the previous section, the sphaleron transition in the
symmetric phase and other elementary interactions are in equilibrium
at $T\sim T_C$, because of slow expansion rate of the universe.
Hence static properties of the EWPT, such as the transition temperature,
the order of the transition and the latent heat when the transition is 
first order, are described well by the equilibrium statistical 
mechanics.
For this purpose, we calculate the effective potential,
which is the free energy density, as a function of the order parameters
and temperature.
In the case of a first-order phase transition, the effective potential
has the form depicted in Fig.~\ref{fig:2}, in which
\begin{equation}
 \varphi_C \equiv \lim_{T\uparrow T_C}\varphi(T)\not= 0 
        \label{eq:def-phiC}
\end{equation}
characterizes the order of the transition, with $\varphi(T)$ being
the order parameter.\par
%
%%%%%%% Fig.2 %%%%%%%%
The first attempt to evaluate the effective potential was based on
the loop expansion in finite-temperature field 
theory\cite{Kirzhnits,Dolan,Sher}.\  
The effective potential in the minimal standard model (MSM),
at the one-loop level, is given by
\begin{equation}
 V_{\rm eff}(\varphi;T)=V_{\rm tree}(\varphi) + V^{(1)}(\varphi;T),
               \label{eq:Veff-SM1}
\end{equation}
where $V_{\rm tree}(\varphi)$ is the tree-level potential,
\begin{equation}
 V_{\rm tree}(\varphi) = -\half\mu_0^2\varphi^2+{{\lambda_0}\over4}\varphi^4,
      \label{eq:Vtree-SM}
\end{equation}
$V^{(1)}(\varphi;T)$ is the one-loop contribution
\begin{equation}
 V^{(1)}(\varphi;T) = -{i\over2}\sum_A c_A\int_k\log\det\left[
                       i{\cal D}_A^{-1}(k;\varphi)\right]
      \label{eq:Vone-loop-SM}
\end{equation}
and the order parameter $\varphi$ is introduced as the expectation 
value of the Higgs field as
\begin{equation}
   \expecv{\Phi(x)} = {1\over{\sqrt{2}}}\pmatrix{0\cr \varphi\cr}.
       \label{eq:def-phi-SM}
\end{equation}
Here $\mu_0^2$ and $\lambda_0$ are the bare parameters, which will be
determined once one prescribes the renormalization. 
In(\ref{eq:Vone-loop-SM}), the subscript $A$ runs over all the species
including the Faddeev-Popov ghosts, $c_A$ denotes the degrees of freedom
of each particle and its statistics, that is, $c_A>0$ for bosons and $c_A<0$
for fermions, and ${\cal D}^{-1}$ is the inverse propagator, which
is in general a matrix with Lorentz and internal symmetry indices 
and depends on $\varphi$ through its mass. The integration over
the momentum should be understood as
\begin{equation}
 \int_k\equiv iT\!\sum_{n=-\infty}^{\infty}\int{{d^3{\mib k}}\over{(2\pi)^3}}
 \qquad \mbox{with  }
   k^0=\omega_n=\left\{\begin{array}{ll}
                       2n\pi T  &\mbox{ for bosons},\\
                       (2n+1)\pi T  &\mbox{ for fermions}.
                       \end{array} \right.
             \label{eq:FT-integral}
\end{equation}
For example, $c_W=2$ and 
$i{{\cal D}^{-1}}_W^{\mu\nu}(k;\varphi)=(-k^2+m_W^2(\varphi))\eta^{\mu\nu}+
 (1-\xi^{-1})k^\mu k^\nu$ for $W$ boson in the $R_\xi$-gauge, where
$m_W(\varphi)=\half g\varphi$, $c_f=-2$ and 
$i{\cal D}_f^{-1}(k;\varphi)=\kslash-m_f(\varphi)$ for a Dirac fermion
with $m_f(\varphi)= y_f\varphi/\sqrt{2}$.
The divergences in (\ref{eq:Vone-loop-SM}) are absorbed in the bare 
parameters by the renormalization at zero temperature.
For the Higgs boson, $m^2_H(\varphi) = 3\lambda\varphi^2-\mu^2$,
so that $V^{(1)}(\varphi;T)$ becomes complex for small 
$\varphi$\cite{Dolan}.\   Since the free energy is originally a real
quantity, this pathology will be caused by the loop expansion, which is
invalid for small $\varphi$, and is expected to be cured if one takes
the higher order contributions into account.
At this point, we neglect the contributions from the Higgs and higher
orders and discuss the structure of $V_{\rm eff}(\varphi;T)$.
Now we consider only $W$ and $Z$ bosons and top quarks, since 
the contributions from the other fermions are negligible. 
Then the one-loop effective potential is
\begin{equation}
 V_{\rm eff}(\varphi;T) = V_0(\varphi) + {\bar V}(\varphi;T),
        \label{eq:Veff-SM2}
\end{equation}
where
\begin{eqnarray}
 V_0(\varphi)&=&
  -\half\mu^2\varphi^2+{\lambda\over4}\varphi^4+2Bv_0^2\varphi^2
  + B\varphi^4\left[\log\left({{\varphi^2}\over{v_0^2}}\right)
                    -{3\over2} \right],    \label{eq:Veff-0-SM}\\
 {\bar V}(\varphi;T)&=&
  {{T^4}\over{2\pi^2}}\left[ 6I_B(a_W) + 3I_B(a_Z) - 6I_F(a_t)
                      \right].             \label{eq:Veff-T-SM}
\end{eqnarray}
Here we follow the renormalization convention of Ref.~\cite{DineLeigh},
and
\begin{eqnarray}
 B&=&
  {3\over{64\pi^2v_0^4}}(2m_W^4+m_Z^4-4m_t^4), \label{eq:def-BinVeff}\\
 I_{B,F}(a)&=&
  \int_0^\infty dx\, x^2\log\left(1\mp{ e}^{-\sqrt{x^2+a^2}}
                               \right),        \label{eq:int-inVeff}
\end{eqnarray}
where $v_0=246\mbox{GeV}$ is the minimum of $V_0(\varphi)$ and
$a_A = m_A(\varphi)/T$.\par
For high temperatures $T>m_W,m_Z,m_t$, the integrals $I_{B,F}$ can be
approximated by the high-temperature expansion\cite{Dolan}:
\begin{equation}
 V_{\rm eff}(\varphi;T)\simeq D(T^2-T_0^2)\varphi^2 - ET\varphi^3
                         + {{\lambda_T}\over4}\varphi^4,
                                    \label{eq:Veff-highT-SM}
\end{equation}
where
\begin{eqnarray}
 D&=& {1\over{8v_0^2}}(2m_W^2+m_Z^2+2m_t^2), \label{eq:def-D-SM}\\
 E&=& {1\over{4\pi v_0^3}}(2m_W^3+m_Z^3)\sim 10^{-2},
                                             \label{eq:def-E-SM}\\
 T_0^2&=& {1\over{2D}}(\mu^2-4Bv_0^2),       \label{eq:def-T0-SM}\\
 \lambda_T&=&
  \lambda-{3\over{16\pi^2v_0^4}}\left(
    2m_W^4\log{{m_W^2}\over{\alpha_B T^2}} +
     m_Z^4\log{{m_Z^2}\over{\alpha_B T^2}}   \right.\nonumber\\
  & &\qquad\qquad\qquad \left.     
     -4m_t^4\log{{m_t^2}\over{\alpha_F T^2}} \right)
                                             \label{eq:def-lamT-SM}
\end{eqnarray}
with $\log\alpha_B=2\log\,4\pi-2\gamma_E$,
$\log\alpha_F=2\log\,\pi-2\gamma_E$ and $\gamma_E=0.5772\cdots$ being
the Euler constant.
The effective potentials of other models which exhibit first-order phase
transition will have a form similar to that of (\ref{eq:Veff-highT-SM})
with $E>0$, while each coefficient will be modified appropriately. 
By use of this approximate form, we see that at $T=T_C$, where 
the transition temperature $T_C$ is determined by
\begin{equation}
  T_C^2 = {{T_0^2}\over{1-E^2/(\lambda_{T_C}D)}},   \label{eq:Tc-SM}
\end{equation}
the effective potential has two degenerate minima at $\varphi=0$ and
\begin{equation}
 \varphi_C = {{2ET_C}\over{\lambda_{T_C}}}.       \label{eq:phiC-SM}
\end{equation}
The presence of the $\varphi^3$-term with positive $E$ in the effective
potential is essential for the EWPT to be first order. It originates from
the zero-frequency mode of the bosons in the finite-temperature 
Feynman integral in (\ref{eq:Veff-T-SM}).
This can be seen as follows\cite{DineLeigh}.\  
The zero-frequency contribution to $dV_{\rm eff}/d\varphi$ has
the form of
\begin{equation}
 {{dV_{\rm eff}(\varphi;T)}\over{d\varphi}}\sim
 {{dm^2(\varphi)}\over{d\varphi}}\int{{d^3{\mib k}}\over{(2\pi)^3}}
 {1\over{{\mib k}^2+m^2(\varphi)}}
 = -{1\over{4\pi}}{{dm^2(\varphi)}\over{d\varphi}}\sqrt{m^2(\varphi)},
                               \label{eq:dV/dphi-SM}
\end{equation}
where the divergence which is absorbed by renormalization is dropped.
Since this is proportional to $\varphi^2$, $V_{\rm eff}$ contains a
$\varphi^3$-term upon integration.
Note that within this approximation, the EWPT appears to be first order,
since $V_{\rm eff}$ always has a $\varphi^3$-term whose coefficient has
the correct sign.\par
Once the expectation value of the Higgs at $T_C$ is known, one can
evaluate the sphaleron rate at $T_C$. Requiring that the sphaleron 
process decouples in the broken phase just after the EWPT imposes
an upper bound on the Higgs mass\cite{mH-bound-SM}.\  
This amounts to the condition that the sphaleron rate in the broken
phase (\ref{eq:sph-rate-br}) multiplied by particle density $\sim T^3$
is smaller than the Hubble parameter (\ref{eq:HatTc}), which yields
\begin{equation}
  {{\varphi_C}\over{T_C}} \gtsim 1.      \label{eq:bound-phi/T}
\end{equation}
This is converted to an upper bound on $\lambda$. Since the Higgs mass
is given by $m_H={\sqrt 2\lambda}v_0$ at the tree level, 
$m_H$ is bounded as
\begin{equation}
  m_H \ltsim 46\mbox{GeV}.               \label{eq:bound-mH-SM1}
\end{equation}
This is obviously inconsistent with the present lower bound
$m_H>58.4\mbox{GeV}$\cite{PDG}.\  
Although this upper bound may be a crude value, the decoupling of 
sphaleron process in general will require the Higgs be lighter than
some bound. This upper bound will be relaxed if one extends the model
to include extra bosons, which effectively enhance the 
$\varphi^3$-term.\par
Among these extended models are those with two Higgs doublets,
including the minimal supersymmetric standard model (MSSM).
We studied the massless two-Higgs-doublet model, in which
the gauge symmetry is broken by the Coleman-Weinberg mechanism,
without recourse to the high-temperature expansion.
Our result shows that for the entire range of the parameters we 
considered,
the EWPT is first order, the sphaleron decoupling condition is
satisfied and the lightest Higgs scalar is as heavy as 
$63\mbox{GeV}$\cite{FKT}.\  
The more general two-Higgs-doublet model was studied by randomly
scanning the vast parameter space\cite{Davis}.\  
For the parameters consistent with present experiments and allowing
perturbation, the lightest Higgs scalar has mass
$45\ltsim m_h\ltsim190\mbox{GeV}$.
The upper bound on the Higgs mass in the MSSM was evaluated for
various values of the soft supersymmetry-breaking 
parameters and was found to be near the experimental lower 
bound\cite{Myint}.\  We note that all these results are based on 
the one-loop effective potential and the latter two works used
the high-temperature expansion.
In general, for these extended models, we have more chances to
satisfy the bound on the Higgs mass consistent with experiments and
to obtain first-order EWPT.\par
Here we shall comment on higher-order studies of the effective 
potential. The most dominant contributions are the daisy type diagrams,
and summing up these gives a well-defined effective potential\cite{Dolan}.\  
Since these diagrams correspond to infinite-order mass 
corrections, some authors evaluate the `improved' effective potential
simply by replacing $m^2$ of the particle in the loop with
$m^2+\Pi(\varphi)$ in the one-loop effective potential, 
where $\Pi(\varphi)$ is the one-loop polarization.
Although this prescription may somehow cure the pathology of complex
effective potential for smaller $\varphi$, it does not yield correct
answer\cite{DineLeigh}.\  The correct method gives a smaller value of
$E$ in (\ref{eq:Veff-highT-SM}), which means a weaker first-order
phase transition. Some two-loop calculations suggest the opposite
situation. The two-loop corrections to the ring-improved one-loop
effective potential were found to raise the lower bound derived from 
the sphaleron decoupling condition, while it still rules out the 
MSM\cite{ArnoldEspinosa}.\footnote{This calculation was carried out
in the Landau gauge. As for the gauge-dependence of
the effective potential, see Ref.~\cite{Kripfganz}.}\  
The two-loop effective potential was shown to increase the strength 
of the first-order EWPT\cite{FodorHebecker}.\  
These two-loop corrections are so large that we still need more reliable
evaluation for a definitive conclusion.\par
Besides these analytic investigations, the static properties
of the EWPT have been studied numerically by the lattice 
simulations\cite{MCreview}.\  
Since at present the electroweak theory cannot be put on the lattice
completely, its subsystem, $SU(2)$ gauge-Higgs system with one Higgs
doublet, has been taken, and two methods have been applied to the simulation.
At high temperatures, fermions attain effective masses of $O(T)$ because
of their antiperiodicity along the imaginary time direction, even if
they are massless at zero temperature.
This fact might justify this approximation neglecting all the fermions
in the model at high temperatures.
One of the simulations is the standard one on the periodic euclidean
four-dimensional lattice. The other is that of the effective 
three-dimensional theory, which is obtained by taking the high-temperature
limit. In this limit, the euclidean time period, $1/T$, is smaller
and the system is reduced to that of the fields, which are zero modes
of the Matsubara frequency in the original system.
That is, the effective three-dimensional system is composed of an
$SU(2)$ gauge field with one Higgs doublet and one Higgs triplet, 
which is the remnant of the time-component of the original gauge field.
Correspondingly, the parameters come to have temperature-dependence,
which are determined by evaluating finite temperature Green's 
functions in both the original and the reduced theory and by matching 
them\cite{Kajantie}.\  
Because of Elitzur's theorem, no local symmetry is broken
on the lattice, so that $\expecv{\phi}=0$\cite{Elitzur}.\  
The measured quantities are expectation values of the gauge invariant
operators such as $\expecv{\phi^\dagger\phi}$, its jump, transition
temperature $T_C$, latent heat and surface tension.
The latent heat is obtained by the derivative with respect to 
temperature of the difference of the free energy between the two phases,
just as the continuum theory. There are three methods to measure
the surface tension between the two phases.
As for the comparison of the results of these simulations with the 
continuum analysis, see Ref.~\cite{Jansen}.  
According to Ref.~\cite{Jansen}, the numerical results coincide with 
those of the continuum two-loop perturbation theory, up to the Higgs masses
of about $70\mbox{GeV}$. For example, the three-dimensional reduced
theory gives, with both perturbative and numerical methods, 
$T_C\simeq93\mbox{GeV}$ and $\varphi_C/T_C\simeq1.8$ for $m_H=35\mbox{GeV}$,
$T_C\simeq140\mbox{GeV}$ and $\varphi_C/T_C\simeq0.7$ for $m_H=60\mbox{GeV}$,
$T_C\simeq155\mbox{GeV}$ and $\varphi_C/T_C\simeq0.6$ for $m_H=70\mbox{GeV}$.
It also yields the result that if we denote the surface tension by $\sigma$,
$\sigma/T_C^3\simeq 0.84$ for $m_H=18\mbox{GeV}$,
$\sigma/T_C^3\simeq 0.065$ for $m_H=35\mbox{GeV}$ and
$\sigma/T_C^3\simeq 0.008$ for $m_H=49\mbox{GeV}$ by the 
four-dimensional simulation.
From these results, the EWPT of the MSM is first 
order for $m_H\ltsim 70\mbox{GeV}$. The strength of the transition
rapidly decreases as $m_H$ increases.
For $m_H\ge60\mbox{GeV}$, (\ref{eq:bound-phi/T}) is not satisfied,
but the upper bound on $m_H$ might be weakened since temperature
in the broken phase is lower than $T_C$ because of supercooling.\par
These quantities compared between analytic and numerical methods
are those derived from the effective potential but not itself.
One can, in principle, calculate the effective potential by 
numerical methods, but its form is always convex\cite{Sher}\  
so that it does not give quantities such as the height and
width of the barrier of the effective potential dividing the two 
phases. 
Our aim is to obtain a `classical' effective potential
which offers information about the first-order phase transition.
Such an effective potential should reproduce the results
of the lattice simulations and have a shape like Fig.~\ref{fig:2}.\par
As we shall see in the next section, to have an efficient $CP$ violation,
extension of the Higgs sector would be needed.
Further, such extension would open the chance to fulfill the condition
(\ref{eq:bound-phi/T}) within the present Higgs mass bound.
Hence both analytic and numerical studies of such models would
be desired.
\subsection{Dynamics of the phase transition}
To have nonzero BAU, we need a nonequilibrium state,
which is realized by expanding bubbles created at the first-order
EWPT. Whether the EWPT is a first-order transition accompanying
nucleation and consecutive growth of bubbles of the broken phase
in the symmetric phase will be determined by the form of the 
effective potential. If the EWPT proceeds with bubble nucleation and
growth, the width and velocity of the bubble wall are essential to
evaluate the baryon number generated.
The nucleation rate of the bubbles is also crucial, since
if their nucleation dominates over their growth, the total 
region which experiences nonequilibrium conditions will decrease.\par
A crude picture of how the EWPT proceeds can be gained as follows.
Suppose that the effective potential at temperatures around $T_C$
is known. The nucleation rate per unit time and volume is given
by\cite{Langer}
\begin{equation}
 I(T) = I_0 { e}^{-\Delta F(T)/T},       \label{eq:nucl-rate}
\end{equation}
where the prefactor $I_0$ is determined by temperature, viscosity
and correlation length in the symmetric phase and $\Delta F$ is
the change in the free energy caused by formation of a bubble.
As long as the thin-wall approximation is valid, $\Delta F$ is
approximately given by
\begin{equation}
  \Delta F(T) = {{4\pi}\over3}r^3[p_s(T)-p_b(T)]+4\pi r^2\sigma,
         \label{eq:deltaF}
\end{equation}
where $p_{s(b)}$ is the pressure in the symmetric (broken) phase and
is given by
\begin{equation}
  p_s(T) = - V_{\rm eff}(0;T),\qquad
  p_b(T) = - V_{\rm eff}(\varphi(T);T).     \label{eq:pressure}
\end{equation}
In general, $p_s(T) < p_b(T)$ because of supercooling.
$\sigma$ is the surface energy density given by 
$\sigma\simeq\int dz (d\varphi/dz)^2$, with $z$ being the coordinate
perpendicular to the bubble wall. 
For smaller radius $r$, the second term on the right-hand side of
(\ref{eq:deltaF}) dominates so that the surface tension shrinks the
bubbles. For larger $r$, the first term dominates and the bubble 
expands to lower the free energy. The critical bubble has a boundary
radius between these bubbles, whose radius is
\begin{equation}
 r_*(T) = {{2\sigma}\over{p_b(T)-p_s(T)}}. \label{eq:critical-bubble}
\end{equation}
With the free energy $V_{\rm eff}(\varphi;T)$ known, one can calculate
various thermodynamical quantities. For example, the entropy density 
is given by $s=-\del V_{\rm eff}/\del T$ and the energy density is
$\rho = V_{\rm eff}-Ts$. The latent heat is the difference of the energy 
density between the two phases. How the phase transition proceeds may
be characterized by the fraction of space which has been converted to 
the broken phase, $f(t)$. It obeys the integral equation
\begin{equation}
 f(t)=\int_{t_C}^t dt^\prime\, I(T(t^\prime))[1-f(t^\prime)]V(t^\prime,t),
                                          \label{eq:brokenfraction}
\end{equation}
where $V(t^\prime,t)$ is the volume of a bubble at $t$ which was 
nucleated at $t^\prime<t$.
Here, the time $t$ and temperature $T$ are related by the scale factor
of the universe $R$ at $t$, since $\rho\propto R^{-4}$ for the 
radiation-dominated universe and $\rho=(\pi^2/30)g_*T^4$. 
(See Appendix A.) $V(t^\prime,t)$ is taken as
\begin{equation}
  V(t^\prime,t)
  ={{4\pi}\over3}\left[r_*(T(t^\prime))
          +\int_{t^\prime}^t dt^{\prime\prime}\,v(T(t^{\prime\prime}))
                 \right]^3,       \label{eq:bubble-volume}
\end{equation}                           
where $v(T)$ is the wall velocity. This simple approach does not take
into account interactions and fluctuations of the bubbles.
The wall velocity and thickness are estimated by solving dynamical equation
for $\varphi(x)$ with friction\cite{Liu,Moore}.\  
In Ref.\cite{Carrington}, (\ref{eq:brokenfraction}) is numerically
analyzed by use of the wall velocity given in Ref.\cite{Liu}\  
with the one-loop improved effective potential of the MSM
for $m_H=60\mbox{GeV}$ and $m_t=120\mbox{GeV}$.
The results show that if we measure the time after the temperature reached
$T_C$,
(1) at $6.5\times10^{-14}\mbox{sec}$,
bubbles began to nucleate,\footnote{At this temperature, the characteristic
time scale of the electroweak processes is $O(10^{-26})\mbox{sec}$.}
(2) at $6.87\times10^{-14}\mbox{sec}$, the nucleation was turned off 
and then only about $10\% $ of the universe had been converted to the 
broken phase, then
(3) the remaining $90\%$ of the universe is converted by the 
bubble growth with almost constant velocity of about $0.8$, and
(4) the transition is completed at $7.05\times10^{-14}\mbox{sec}$.
Further if we denote the temperature at which the bubble nucleation 
begins by $T_N$, $(T_C-T_N)/T_C\simeq 2.5\times10^{-4}$, which means
the EWPT shows small supercooling.\par
All these results are based on the effective potential of the MSM
at the one-loop level or the improved version of it.
As mentioned, the above estimation neglects the interactions of the 
bubbles and various fluctuations.
For weak first-order transitions, thermal fluctuations might affect
the dynamics of the transition. Then the barrier dividing
the symmetric and broken phases of $V_{\rm eff}$ is so low that the two
phases may be mixed by thermal fluctuations.
In such a case, the EWPT might proceed not with nucleation of the 
critical bubbles but with that of the subcritical 
bubbles\cite{subc-bubble}\  so that the simple analysis above may no 
longer be valid.\par
The numerical simulations as well as the two-loop effective potential
suggest a stronger first-order transition in the MSM than the one-loop
result. Extensions of the standard model with
more Higgs scalars may also lead to a stronger transition, since
the upper bound on the Higgs mass is raised somehow and since this also 
implies that $\varphi_C$ is larger for a given Higgs mass.
In fact, our work on the two-Higgs-doublet model supports this fact and
thinner bubble walls\cite{FKT}.
%
%
%
% 
%-------------
\section{Mechanism of the electroweak baryogenesis}
\begin{figure}
 \epsfxsize = 8cm
 \centerline{\epsfbox{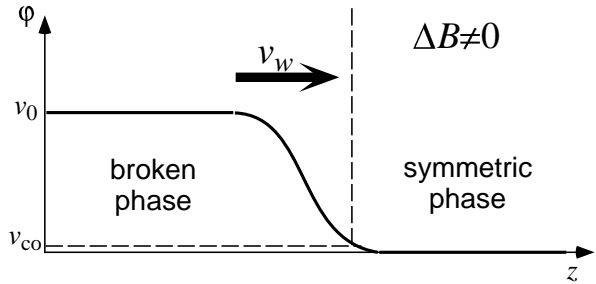}}
 \caption{Profile of the expanding bubble wall. In the region right 
 to the vertical dashed line, the sphaleron process is effective.
 $v_w$ is the wall velocity.}
 \label{fig:3}
\end{figure}

If the sphaleron process is in equilibrium in the symmetric phase and
decouples in the broken phase at the EWPT, it is out of equilibrium
around the moving bubble wall dividing the two phases.
In this section, we review how the baryon number is generated by use of 
this nonequilibrium situation.
Around the expanding bubble wall, the expectation value of the Higgs 
field is schematically depicted in Fig.~\ref{fig:3}, where $\varphi$ 
represents the absolute value of the Higgs scalar. $v_0$ is that in the 
broken phase and $v_{co}$ is the value above which the sphaleron 
process decouples.
%
%%%%%%% Fig.3 %%%%%%%%
There are two mechanisms of the electroweak baryogenesis;
classical or adiabatic mechanism, which is called `spontaneous
baryogenesis' and quantum mechanical one,
which is also called `charge transport scenario'.
Although these two mechanisms may be responsible for 
baryogenesis at the same time, the former is effective for a thick
bubble wall while the latter is effective for a thin wall, 
as we shall see below.
Further both require extensions of the Higgs sector in the MSM, 
except for the scenario proposed by Farrar and Shaposhnikov\cite{Farrar}.\  
The spontaneous scenario was proposed earlier than the charge transport
mechanism.
It, however, was pointed out that it would generate too small a number
of baryons in its original form. After the charge transport scenario
was proposed, the classical mechanism was revived by introducing a
nonlocal effect, which is an advantage of the charge transport scenario.
\subsection{Charge transport scenario}
\begin{figure}
 \epsfxsize = 8cm
 \centerline{\epsfbox{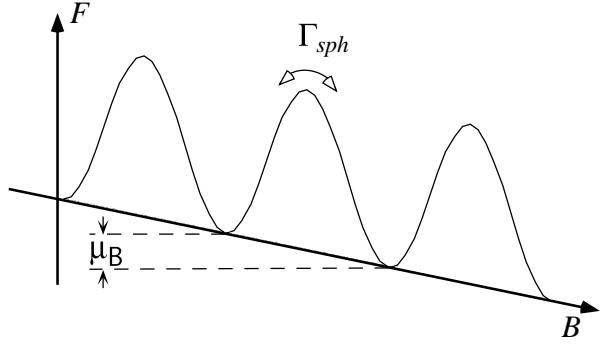}}
 \caption{The free energy as a function of the baryon number is 
 biased and a nonequilibrium stationary state is realized.}
 \label{fig:4}
\end{figure}

Fermions interact with the bubble wall through the Yukawa
couplings, which in general contain $CP$ violation. If the $CP$ violation
is not constant over spacetime, it cannot be rotated out by a biunitary
transformation so that there remains a physical effect.
This $CP$ violation discriminates between the interaction of the fermions and
that of the antifermions with the bubble wall. That is, the reflection
rate of the fermions does not coincide with that of antifermions so 
that the net current of some quantum number flows into the symmetric 
phase region, where the sphaleron process is effective.
There the previous equilibrium state will be forced to shift to new
state, which will contain baryon number excess.
One can see how this occurs by solving the kinetic equations,
which will contain various interactions of different time scales.
Such equations are in general difficult to solve.
When the bubble wall moves with constant velocity, the flux will 
bias the free energy along the baryon number to realize a stationary
nonequilibrium state, as depicted in Fig.~\ref{fig:4}.\  
%
%%%%%%% Fig.4 %%%%%%%%
In this case, the constant flow of the flux will induce a nonzero
chemical potential for the baryon number. Then the steady state
produces the baryon number according to the equation
\begin{equation}
  {\dot n}_B\simeq - {{\mu_B\Gamma_{\rm sph}}\over T}, 
        \label{eq:detailedbalance}
\end{equation}
where $\Gamma_{\rm sph}$ is the sphaleron rate in the symmetric phase
(\ref{eq:sph-rate-sym}). For the derivation of this equation, see
Appendix B.\par
Cohen, Kaplan and Nelson related the flux flowing into the symmetric 
phase to the generated baryon number as follows\cite{NKC}.\  
Suppose that a left(right)-handed fermion of species $i$ with charge
$Q_{L(R)}^i$ is incident from the symmetric phase region.
The expectation value of the injected charge in the symmetric phase,
which has been brought by the reflection and transmission of the 
fermions, is given by
\begin{eqnarray}
 \Delta Q_i^s &=&
 [(Q_R^i-Q_L^i) R^s_{L\rightarrow R} 
     +(-Q_L^i+Q_R^i){\bar R}^s_{L\rightarrow R}   \nonumber\\
 & &
  +(-Q_L^i)(T^s_{L\rightarrow L}+T^s_{L\rightarrow R})
  -(-Q_R^i)({\bar T}^s_{R\rightarrow L}+{\bar T}^s_{R\rightarrow R})]
    f^s_{Li}                                      \nonumber\\
 & &
  +[(Q_L^i-Q_R^i)R^s_{R\rightarrow L}
      +(-Q_R^i+Q_L^i){\bar R}^s_{L\rightarrow R}  \nonumber\\
 & &
  +(-Q_R^i)(T^s_{R\rightarrow L}+T^s_{R\rightarrow R})
  -(-Q_L^i)({\bar T}^s_{L\rightarrow L}+{\bar T}^s_{L\rightarrow R})]
    f^s_{Ri},                   \label{eq:charge-from-s}
\end{eqnarray}
where $R^s_{R\rightarrow L}$ (${\bar R}^s_{R\rightarrow L}$) is the 
reflection coefficient for the right-handed fermion (antifermion)
incident from the symmetric phase region and reflected as the 
left-handed one, $T$'s are the transmission coefficients and
$f^s_{L(R)}$ is the left(right)-handed fermion flux density in the 
symmetric phase, which will be given below.
Here we have used ${\bar f}^s_L=f^s_R$.
In a similar manner, the expectation value of the change of the charge
brought by the transmission of the fermions incident from the broken 
phase region is
\begin{eqnarray}
 \Delta Q_i^b &=&
 Q_L^i(T^b_{L\rightarrow L}f^b_{Li}+T^b_{R\rightarrow L}f^b_{Ri})+
 Q_R^i(T^b_{L\rightarrow R}f^b_{Li}+T^b_{R\rightarrow R}f^b_{Ri})
                                  \label{eq:charge-from-b}\\
 & &
 +(-Q_L^i)({\bar T}^b_{R\rightarrow L}f^b_{Li}+
             {\bar T}^b_{L\rightarrow L}f^b_{Ri})
 +(-Q_R^i)({\bar T}^b_{R\rightarrow R}f^b_{Li}+
             {\bar T}^b_{L\rightarrow R}f^b_{Ri}). 
                                  \nonumber
\end{eqnarray}
By use of the unitarity
\begin{equation}
 R^s_{L\rightarrow R}+T^s_{L\rightarrow L}+T^s_{L\rightarrow R}=1,
 \qquad\mbox{etc.}   \label{eq:unitarity}
\end{equation}
and the reciprocity\cite{FKOTTa}
\begin{equation}
 T^s_{R\rightarrow L}+T^s_{R\rightarrow R}=
 T^b_{L\rightarrow L}+T^b_{R\rightarrow L},
 \qquad\mbox{etc.}   \label{eq:reciprocity}
\end{equation}
the total expectation value of the change is
\begin{equation}
 \Delta Q^s_i+\Delta Q^b_i = (Q_L^i-Q_R^i)(f^s_i - f^b_i)\Delta R,
                      \label{eq:total-change}
\end{equation}
where
\begin{equation}
  \Delta R \equiv R^s_{R\rightarrow L} - {\bar R}^s_{R\rightarrow L},
                      \label{eq:def-deltaR}
\end{equation}
and we have put $f^{s(b)}_{iL}=f^{s(b)}_{iR}\equiv f^{s(b)}_i$.
The total charge injected into the symmetric phase can be obtained by
integrating this in the rest frame of the bubble wall and going back
to the rest frame of the medium,
\begin{equation}
 F^i_Q={{Q_L^i-Q_R^i}\over{4\pi^2\gamma}}
       \int_{m_0}^\infty dp_L \int_0^\infty dp_T\,p_T
       \left[ f_i^s(p_L,p_T)-f_i^b(-p_L,p_T) \right]
       \Delta R({{m_0}\over a},{{p_L}\over a}). \label{eq:def-chiral-flux}
\end{equation} 
Here
\begin{eqnarray}
 f_i^s(p_L,p_T)&=&
 {{p_L}\over E}{1\over{\exp[\gamma(E-v_w p_L)/T]+1}},  \nonumber\\
 f_i^b(-p_L,p_T)&=&
 {{p_L}\over E}{1\over{\exp[\gamma(E+v_w\sqrt{p_L^2-m_0^2})/T]+1}},
           \label{eq:def-flux-density}
\end{eqnarray}
are the fermion flux densities in the symmetric and broken phase,
respectively, $m_0$ is the fermion mass in the broken phase,
$v_w$ is the wall velocity, $\gamma=1/\sqrt{1-v_w^2}$,
$E=\sqrt{p_L^2+p_T^2}$, $p_T$ being the transverse momentum with 
respect to the bubble wall.
If we calculate $\Delta R$ with zero-temperature Dirac equation
in the presence of the $CP$-violating potential, it is a function of
$m_0/a$ and $p_L/a$, where $a^{-1}$ is the wall width.\par
$\Delta R$ contains the information of $CP$ violation for the 
model with which we are concerned. In the MSM, $CP$ violation 
enters only through the Kobayashi-Maskawa (KM) matrix in the 
charged-current interactions.
Farrar and Shaposhnikov proposed that even in the MSM, the interactions
of the quarks with the plasma of the gauge bosons and Higgs bosons,
which modify the dispersion relations of the quasiparticles,
may yield significant baryon number\cite{Farrar}.\  
The effect of the KM matrix comes into the dispersion relations of
$O(\alpha_W)$ and there are claims that QCD damping effects, whose
scale is shorter than that of other interactions, may bring quantum 
decoherence\cite{Gavela}.\  
Hence to obtain enough BAU only via electroweak baryogenesis,
we would have to employ an extra source of $CP$ violation, such
as that in the Higgs sector in the two-Higgs-double model.
The $CP$ violation in the Higgs sector affects the fermion propagation
around the bubble wall at the tree level, so it was the first model 
considered in the charge transport scenario\cite{NKC}.\  
As for the quantum number injected in the symmetric phase, we must
take that conserved there and satisfying $Q_L-Q_R\not=0$.
One such charge is the weak hypercharge $Y$.
Now let us see how the injected hypercharge biases baryon 
number.\par
As we stated above, the change of state caused by the injection of 
hypercharge would be described by the kinetic equations.
Here we assume that the bubble wall is macroscopic and moves with
almost constant velocity, and that the state shifts to a new steady 
state. Such a situation would be described by the chemical 
potentials for all the particles in equilibrium.
This simple picture would be valid in the region far from the wall and
in the case that the elementary processes are fast enough to realize
a new stationary state.
For simplicity, we consider only the third generation and the case 
that all the particles had been in equilibrium with $n_B=n_L=0$ through
the gauge and Yukawa interactions before the injection of the hypercharge.
Now we introduce the chemical potentials
$$
     \mu_{t_L}, \mu_{b_L},\mu_{t_R}, \mu_{b_R}, \mu_{\tau_L}, \mu_{\nu_\tau},
     \mu_{\tau_R},
     \mu_W \mbox{ for $W^-$},
     \mu_0 \mbox{ for $\phi^0$}, \mu_-\mbox{ for $\phi_-$}.
$$
When the charged-current interaction is in chemical equilibrium,
\begin{equation}
  \mu_W=\mu_0+\mu_-=-\mu_{t_L}+\mu_{b_L}=-\mu_{\nu_\tau}+\mu_{\tau_L},
                          \label{eq:chemeq-W}
\end{equation}
and when the Yukawa interaction is in chemical equilibrium,
\begin{equation}
 \mu_0=-\mu_{t_L}+\mu_{t_R}=-\mu_{b_L}+\mu_{b_R}=-\mu_{\tau_L}+\mu_{\tau_R}.
                          \label{eq:chemeq-Higgs}
\end{equation}
No further new equations are derived if the other elementary 
processes are in equilibrium. We also introduce the chemical 
potentials for quantum numbers which are conserved or almost conserved
because of the slow processes:
$$
  \mu_{B-L},\quad \mu_Y,\quad \mu_{I_3};\quad \mu_B.
$$
Here we assume that the sphaleron process is out of equilibrium.
Otherwise, $\mu_B$ vanishes so that the baryon number is not generated.
The chemical potentials of the particles are expressed by these as
\begin{eqnarray}
 \mu_{t_L(b_L)}&=&{1\over3}\mu_B+{1\over3}\mu_{B-L}+{1\over6}\mu_Y
                  +(-){1\over2}\mu_{I_3},                \nonumber\\
 \mu_{t_R}     &=&{1\over3}\mu_B+{1\over3}\mu_{B-L}+{2\over3}\mu_Y,\nonumber\\
 \mu_{b_R}     &=&{1\over3}\mu_B+{1\over3}\mu_{B-L}-{1\over3}\mu_Y,   
                                   \nonumber\\
 \mu_{\tau_L(\nu_\tau)}
               &=&-\mu_{B-L}-{1\over2}\mu_Y+(-){1\over2}\mu_{I_3}, \nonumber\\
 \mu_{0(-)}    &=& +(-){1\over2}\mu_Y-{1\over2}\mu_{I_3},          \nonumber\\
 \mu_W         &=& -\mu_{I_3}.      \label{eq:particle-chem}
\end{eqnarray}
Because of (\ref{eq:n-mu}), the baryon and lepton number densities are
given by
\begin{eqnarray}
 n_B&=&3\cdot{{T^2}\over6}(\mu_{t_L}+\mu_{t_R}+\mu_{b_L}+\mu_{b_R})
     ={{T^2}\over3}(2\mu_B+2\mu_{B-L}+\mu_Y),   \label{eq:B-density}\\
 n_L&=&{{T^2}\over6}(\mu_{\nu_\tau}+\mu_{\tau_L}+\mu_{\tau_R})
     ={{T^2}\over6}(-3\mu_{B-L}-2\mu_Y).        \label{eq:L-density}
\end{eqnarray}
If $n_B=n_L=0$ before the injection of the hypercharge flux,
\begin{equation}
 \mu_{B-L} = -{2\over3}\mu_Y,\qquad \mu_B={1\over6}\mu_Y,
           \label{eq:chempotrel}
\end{equation}
hold because of (\ref{eq:B-density}) and (\ref{eq:L-density}).
Then the hypercharge density, which is expressed by the chemical 
potentials of the particles, can be given, with the help of
(\ref{eq:particle-chem}) and (\ref{eq:chempotrel}), by
\begin{eqnarray}
 {Y\over2}&=&
 {{T^2}\over6}\left\{
  3\left[{1\over6}(\mu_{t_L}+\mu_{b_L})+{2\over3}\mu_{t_R}-{1\over3}\mu_{b_R})
   \right]-{1\over2}(\mu_{\nu_\tau}+\mu_{\tau_L})-\mu_{\tau_R} \right\}
                    \nonumber\\
  & &\qquad
             +{{T^2}\over3}{1\over2}(\mu_0-\mu_-)m          \nonumber\\
          &=&{{T^2}\over6}(m+{5\over3})\mu_Y,    \label{eq:Ybychempot}
\end{eqnarray}
where $m$ is the number of the Higgs doublets. From 
(\ref{eq:chempotrel}) and (\ref{eq:Ybychempot}), we have
\begin{equation}
  \mu_B = {Y\over{2(m+5/3)T^2}}.     \label{eq:muBbyY}
\end{equation}
Thus the injected hypercharge biases the free energy along the 
direction of the baryon number.
By use of (\ref{eq:detailedbalance}), we have the generated baryon 
number density
\begin{equation}
 n_B = -{{\Gamma_{\rm sph}}\over T}\int dt\, \mu_B
     = -{{\Gamma_{\rm sph}}\over{2(m+5/3)T^3}}\int dt\, Y. 
                   \label{eq:nBbyY}
\end{equation}
If we denote the hypercharge density at distance $z$ from the bubble 
wall by $\rho_Y(z)$, the integral of $Y$ is estimated as
\begin{equation}
 \int dt\, Y= \int_{-\infty}^{z/v_w}dt\,\rho_Y(z-v_wt)
            = {1\over{v_w}}\int_0^\infty dz\,\rho_Y(z).  
                  \label{eq:Y-integral}
\end{equation}
Since the hypercharge supplied from the wall would reach a finite
distance when it is caught up with the wall, the last integral above
will be approximated as
\begin{equation}
 {1\over{v_w}}\int_0^\infty dz\,\rho_Y(z)\simeq {{F_Y\tau}\over{v_w}},
            \label{eq:def-tau}
\end{equation}
where $\tau$ is the transport time within which the scattered fermions
are captured by the wall.
Hence the generated baryon asymmetry is
\begin{equation}
 {{n_B}\over s}\simeq{\cal N}{{100}\over{\pi^2g_*}}\cdot\kappa\alpha_W^4
                      \cdot {{F_Y}\over{v_w T^3}}\cdot\tau T, 
                      \label{eq:nB-overs}
\end{equation}
where ${\cal N}$ is a model-dependent constant of $O(1)$.
$\tau$ may be approximated by the mean free time or diffusion time
$D/v_w$ with the diffusion constant $D$ of the charge carrier, 
where $D\simeq1/T$ for quarks and $D\simeq(10^2\sim10^3)/T$ for 
leptons\cite{Joyce1}.\ 
It was shown that if the forward scattering is enhanced, even for the 
top quark, $\tau T\simeq10\sim10^3$ depending on $v_w$, where the 
maximum value is realized at $v_w\simeq1/\sqrt{3}$\cite{NKC}.\  
For this optimal case,
\begin{equation}
 {{n_B}\over s}\simeq 10^{-3} \cdot {{F_Y}\over{v_w T^3}}.
                      \label{eq:opt-BAU}
\end{equation}
Then the hypercharge flux $F_Y/(v_w T^3)\sim O(10^{-7})$ would be
sufficient to explain the present BAU.
For leptons, $\tau T$ is much larger than that of quarks,
since its propagation is not disturbed by strong interactions.
As we shall see below, the hypercharge flux depends not only on
the mass of the carrier but also the wall width and velocity.\par
\begin{figure}
 \epsfxsize = 7cm
 \centerline{\epsfbox{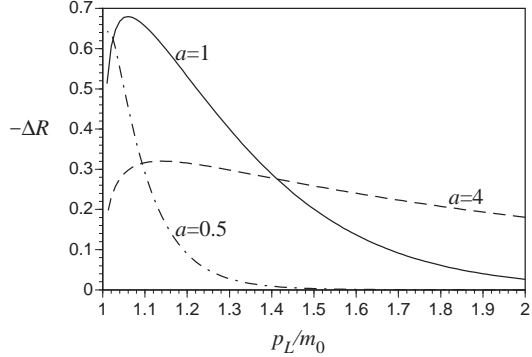}}
 \caption{$\Delta R$ for the potential %(\ref{eq:m-NKC}).
 ({5\raise .5ex\hbox {.}27}). We take $m_0=1$.}
 \label{fig:5}
\end{figure}
\begin{figure}
 \epsfxsize = 7cm
 \centerline{\epsfbox{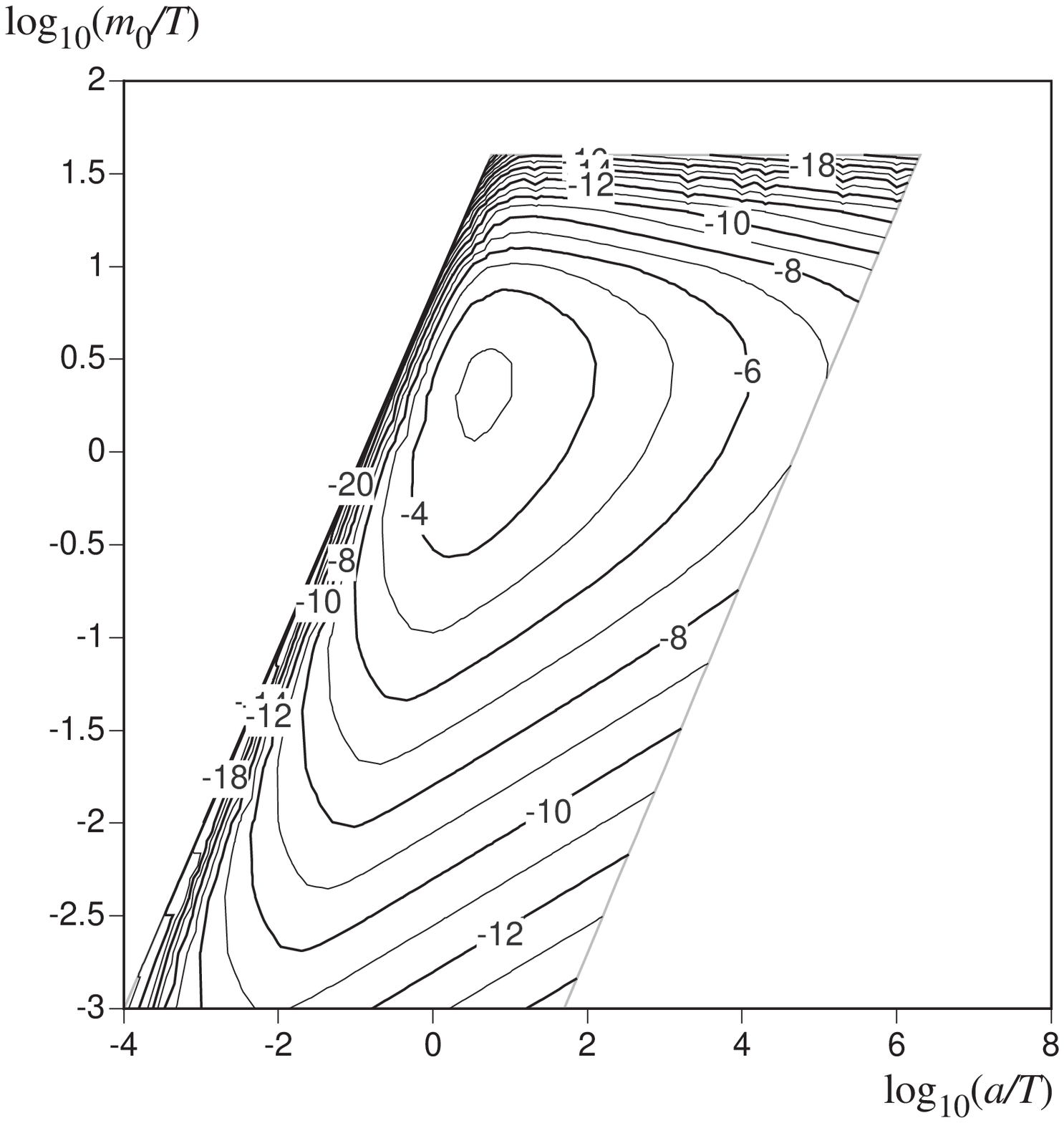}}
 \caption{Contour plot of the chiral charge flux, normalized as
 $\log_{10}\left[-F_Q/(v_w T^3(Q_L-Q_R))\right]$ for the potential
 ({5\raise .5ex\hbox {.}27}). %(\ref{eq:m-NKC}). 
 Here we take $v_w=0.1$ and $T=100\mbox{GeV}$.}
 \label{fig:6}
\end{figure}
\begin{figure}
 \epsfxsize = 7cm
 \centerline{\epsfbox{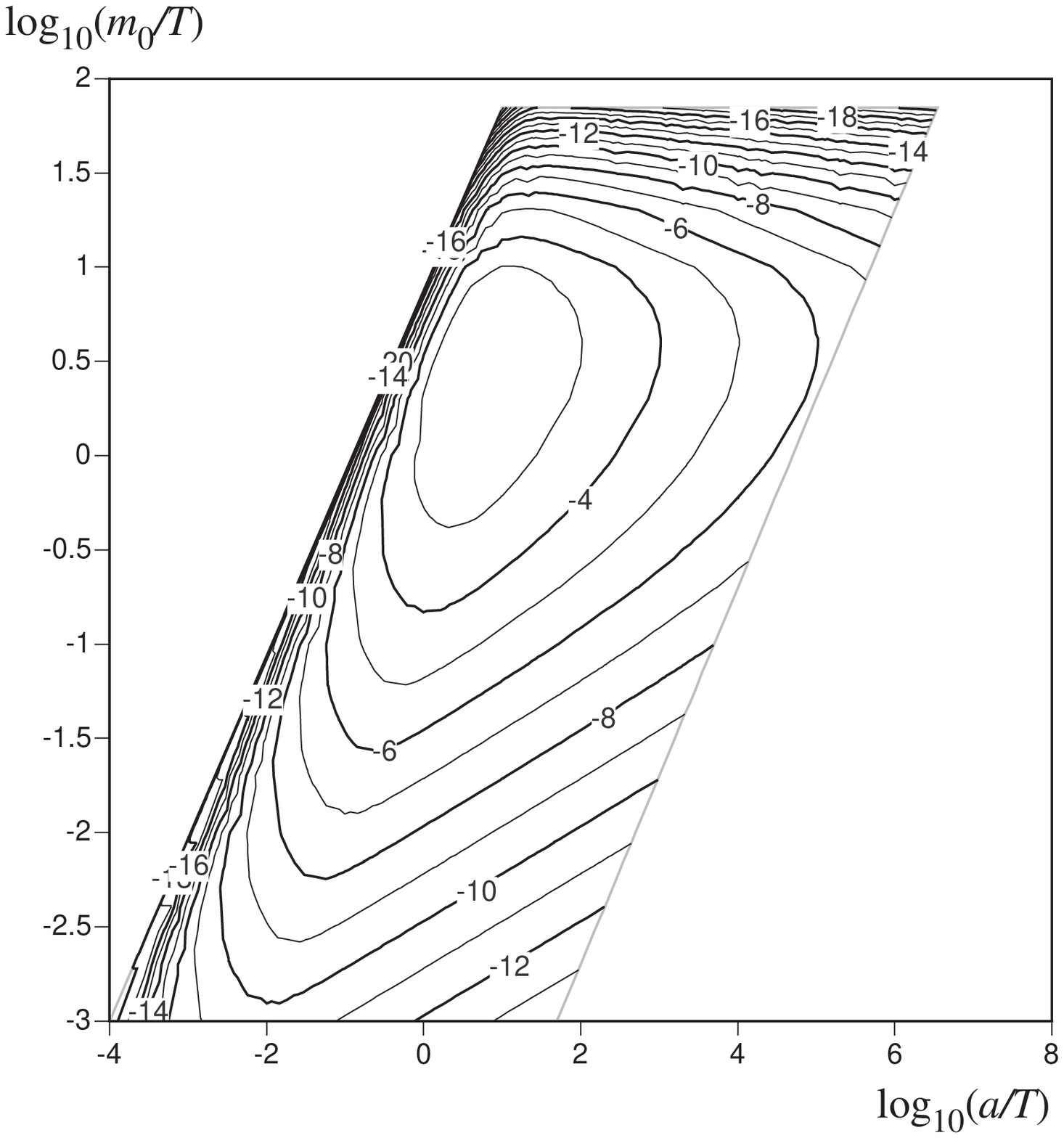}}
 \caption{Contour plot of the chiral charge flux, normalized as
 $\log_{10}\left[-F_Q/(v_w T^3(Q_L-Q_R))\right]$ for the potential
 ({5\raise .5ex\hbox {.}27}). %(\ref{eq:m-NKC}). 
 Here we take $v_w=0.58$ and $T=100\mbox{GeV}$.}
 \label{fig:7}
\end{figure}
\begin{figure}
 \epsfxsize = 7cm
 \centerline{\epsfbox{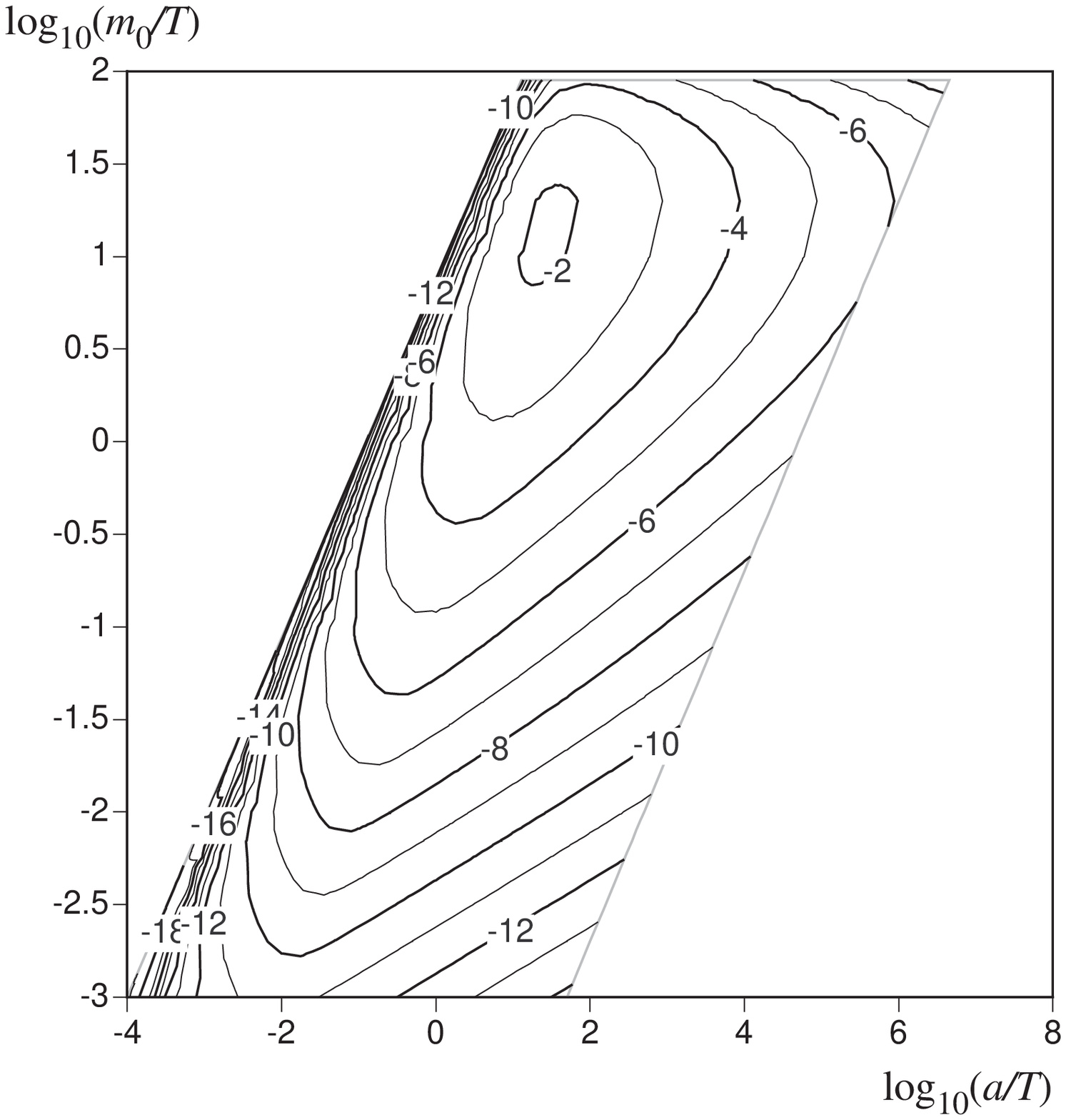}}
 \caption{Contour plot of the chiral charge flux, normalized as
 $\log_{10}\left[-F_Q/(v_w T^3(Q_L-Q_R))\right]$ for the potential
 ({5\raise .5ex\hbox {.}27}). %(\ref{eq:m-NKC}). 
 Here we take $v_w=0.98$ and $T=100\mbox{GeV}$.}
 \label{fig:8}
\end{figure}

Now we evaluate the chiral charge flux in (\ref{eq:def-chiral-flux}).
Here we assume that the simple zero-temperature Dirac equation
describes the propagation of fermions well. This is justified
when the mean free path of the fermion is larger than the bubble wall
width. Since the dynamics of the EWPT in the extended models is not
well established, we treat the wall width and velocity as 
parameters and calculate the flux as a function of them.
The problem is to solve the Dirac equation in the background of the 
bubble wall. The bubble wall is composed of classical gauge and Higgs 
fields. As we shall see in the next section, its profile will
be composed of the Higgs scalar only and the gauge fields are 
pure-gauge type so that we fix the gauge such that the gauge fields 
vanish. Then the Dirac equation to be solved is, for one flavor,
\begin{equation}
 i\dslash\psi(x)-m(x)P_R\psi(x)-m^*(x)P_L\psi(x) = 0,
     \label{eq:Dirac1}
\end{equation}
where $P_R={{1+\gamma_5}\over2}$,$P_L={{1-\gamma_5}\over2}$ and 
$m(x)=-f\expecv{\phi(x)}$ is a complex-valued function of spacetime,
with $f$ being the Yukawa coupling.
When the radius of the bubble is macroscopic and the bubble is
static or moving with a constant velocity, we can regard $m(x)$ as
a static function of only one spatial coordinate:
$$
   m(x) = m(t,{\mib x}) = m(z).
$$
Putting
\begin{equation}
 \psi(x) = { e}^{i\sigma(-Et+{\mib p}_T\cdot{\mib x}_T)},\qquad
 (\sigma=+1\mbox{ or } -1)
  \label{eq:separation}
\end{equation}
(\ref{eq:Dirac1}) is reduced to
\begin{equation}
 \left[ \sigma(\gamma^0E-\gamma_Tp_T)+i\gamma^3\del_z
    -m_R(z) - i\gamma_5 m_I(z) \right] \psi_E({\mib p}_T,z) = 0,
   \label{eq:Dirac2}
\end{equation}
where
\begin{eqnarray*}
 {\mib p}_T&=&(p^1,p^2),\qquad {\mib x}_T=(x^1,x^2),
 \qquad p_T = \absv{{\mib p}_T},\qquad      \\
 \gamma_T p_T &=& \gamma^1 p^1 + \gamma^2 p^2,  \\
 m(z) &=& m_R(z) + i m_I(z). 
\end{eqnarray*}
If we denote $E=E^*\cosh\eta$ and $p_T=E^*\sinh\eta$ with
$E^*=\sqrt{E^2-p_T^2}$, ${\mib p}_T$ in (\ref{eq:Dirac2}) can be
eliminated by the Lorentz transformation
$\psi\mapsto\psi^\prime = { e}^{-\eta\gamma^0\gamma_5}\psi$:
\begin{equation}
 \del_z\psi_E(z)=i\gamma^3\left[-\sigma E^*\gamma^0
                 +m_R(z) + i\gamma_5 m_I(z) \right]\psi_E(z).
	\label{eq:Dirac3}
\end{equation}
This equation was first solved numerically in a finite region
containing the bubble wall\cite{NKC},\  with the following
potential
\begin{equation}
 m(z)=m_0{{1+\tanh(az)}\over2}
      \exp\left(-i\pi{{1-\tanh(az)}\over2}\right). \label{eq:m-NKC}
\end{equation}
The nontrivial $z$-dependence of the phase of $m(z)$ yields $CP$ 
violation in the Dirac equation. If it were independent of $z$,
the phase could be eliminated by redefinition of the wave functions,
so that it would yield no physical effect.
We analyzed (\ref{eq:Dirac3}) by the perturbative 
method\cite{FKOTTa},\  which can be applied when the imaginary part
of the mass function is much less than $m_0$, as well as by
numerical methods\cite{FKOTb},\  which solves (\ref{eq:Dirac3}) in the 
infinite region and gives the results with any precision.
In Fig.~\ref{fig:5}, we show data for $\Delta R$ with the potential
(\ref{eq:m-NKC}), which were obtained numerically\cite{FKOTb}.\  
%
%%%%%%% Fig.5 %%%%%%%%
For higher incident energy, $\absv{\Delta R}$ decreases exponentially.
This is because for $p_L>m_0$, the wave function is obtained 
approximately by the WKB method, which infers that $R_{R\rightarrow L}$
and ${\bar R}_{R\rightarrow L}$ are exponentially small.
$\absv{\Delta R}$ can be $O(1)$ when the wall width is comparable to
the Compton wave length of the carrier, {\rm i.e.}, $m_0/a\sim O(1)$.
Note that larger Yukawa coupling does not always yield larger 
asymmetry in the reflection coefficients.
The chiral charge flux, which is essential to determine the generated
BAU, is evaluated by inserting $\Delta R$ into (\ref{eq:def-chiral-flux}).
We normalize it in a dimensionless form as
\begin{equation}
    {{F_Q}\over{u T^3(Q_L-Q_R)}},        \label{eq:normalized-flux}
\end{equation}
which enters in the baryon asymmetry of (\ref{eq:nB-overs}) or
(\ref{eq:opt-BAU}). The numerical results for the potential
(\ref{eq:m-NKC}) are shown in 
Fig.~\ref{fig:6}, Fig.~\ref{fig:7} and Fig.~\ref{fig:8} for the wall
velocities $v_w=0.1$, $v_w=1/\sqrt{3}\simeq0.58$ and $v_w=0.98$,
respectively, as functions of the carrier mass and the wall width.
Here we take $T=100\mbox{GeV}$.
%
%%%%%%% Fig.6 %%%%%%%%
%
%
%%%%%%% Fig.7 %%%%%%%%
%
%
%%%%%%% Fig.8 %%%%%%%%
These figures suggest that the flux rapidly decreases for a
heavier carrier and thicker wall and the region in which
$-F_Q/(v_w T^3(Q_L-Q_R))>10^{-7}$ becomes broader for larger wall
velocity. The maxima in these figures are realized at
$a\sim T$, for which the wave length of the carrier is comparable
to the wall thickness.
For a thick bubble wall $a\simeq 40/T$, fermions of mass
$0.1T<m_0<T$ can yield a flux larger than $10^{-7}$, while for
a thin wall case $a\simeq 1/T$, fermions of mass $0.03T<m_0<10T$
can yield sufficient flux.\par
To calculate the flux, we assumed the form of the complex mass as
(\ref{eq:m-NKC}). It yields no $CP$ violation in the broken phase so
as to be free from any experimental limitations.
However, the profile of the bubble wall should be determined 
dynamically. This is the subject of the next section, and it will
also affect the other scenario of the electroweak baryogenesis.
\subsection{Spontaneous baryogenesis}
This scenario, in contrast to that in the previous subsection,
is classical and effective for thick bubble walls.
Although this scenario has several variants, the first proposed
may be that by Shaposhnikov\cite{mH-bound-SM},\
That case depends on the existence of the condensation of the
Chern-Simons number, which adds a $CP$-violating term in the effective
action as
\begin{equation}
 \Delta S_{\rm eff} = \mu N_{CS},       \label{eq:CS-condensate}
\end{equation}
where $N_{CS}$ is defined by (\ref{eq:def-CS}).
This scenario fails in the MSM since it can induce at most 
$\mu=10^{-15}$ while $\mu=10^{-4}$ is needed to explain 
the BAU\cite{mH-bound-SM}.\  
It is later elaborated by noting that in the two-Higgs-doublet
extension of the MSM, the effective action takes the form as
\begin{equation}
 \Delta{\cal L}_{\rm eff} = -{{g^2N_f}\over{24\pi^2}}\theta(x)
                         F_{\mu\nu}(x){\tilde F}^{\mu\nu}(x),  
                                    \label{eq:theta-FF}
\end{equation}
where $\theta(x)$ is the relative phase of the two Higgs 
doublets\cite{TurokZadrozny},\  
This term is $CP$-invariant since $\theta(x)$ is $CP$ odd, so that 
it is free from any constraint on $CP$ violation.
A nontrivial evolution of $\theta(x)$ during the EWPT would induce
asymmetry like $\mu$ in (\ref{eq:CS-condensate}), upon integrated
by parts. For high temperatures, the factor in front of the
Chern-Simons number is accompanied by $(m_t/T)^2$ coming from the top
quark loop\cite{McLerran}.\    
Cohen {\rm et al.} proposed a different approach, which utilizes
the bias for the hypercharge in place of the Chern-Simons number,
in the two-Higgs-doublet extension of the MSM\cite{CKN-spont},\  
Their observation was that if the neutral component of the two Higgs 
scalar takes the form
\begin{equation}
 \phi_j^0(x)=\rho_j(x){ e}^{i\theta_j},\qquad(j=1,2)
                   \label{eq:CKN-scalars}
\end{equation}
where only $\phi_1$ is supposed to couple to the fermions,
one can eliminate $\theta_1$ from the Yukawa coupling by an
anomaly-free ($U(1)_Y$) rotation, which changes the fermion kinetic 
terms by
\begin{eqnarray}
 & &2\del_\mu\theta_1(x)\left[{1\over6}{\bar q}_L(x)\gamma^\mu q_L(x)
     +{2\over3}{\bar u}_R(x)\gamma^\mu u_R(x)
     -{1\over3}{\bar d}_R(x)\gamma^\mu d_R(x) \right.  \nonumber\\
 & &\qquad\qquad\qquad\left.
     -{1\over2}{\bar l}_L(x)\gamma^\mu l_L(x)
     -{\bar e}_R(x)\gamma^\mu e_R(x) \right],  \label{eq:trf-thata1}
\end{eqnarray}
where summation over generation is implicit.
During the EWPT, $\theta_1(x)$ will behave nontrivially so that
$\expecv{\dot\theta_1}\not=0$. Then $\expecv{\dot\theta_1}$ will
play the role of `charge potential'.
If this biases the hypercharge in the symmetric phase region,
the same procedure to generate the baryon number as discussed
in the charge transport scenario works here.
They estimated the generated baryon asymmetry as
\begin{equation}
 {{n_B}\over s}\simeq 10^{-8}\Delta\theta, \label{eq:spont-BAU}
\end{equation}
where $\Delta\theta$ is the change of $\theta_1$ during the EWPT.
None of these mechanisms rely on any quantum mechanical scattering,
so that their effects are not weakened for thick bubble walls, in
contrast to the charge transport scenario.\par
It was pointed out that these scenarioes cannot generate sufficient
baryon asymmetry, because of an additional suppression factor in
the charge potential\cite{DineThomas}.\  
Since in the symmetric phase the $CP$ violating angle $\theta_i$
does not affect the dynamics because of $\rho_i=0$, 
the charge potential will be zero there.
The hypercharge current in (\ref{eq:trf-thata1}) is the fermionic
one, and taking correctly the contribution from the scalar parts,
the generated term by the rotation is $\del_\mu\theta_1 j^\mu_Y$ with
the total hypercharge current $j^\mu_Y$, which is conserved in the
symmetric phase. On the other hand, its nonconservation in the 
broken phase leads to, at high temperatures,
\begin{equation}
 \del_\mu\theta_1\cdot j^\mu_Y\propto
    {{m_t}^2\over T^2}F_{\mu\nu}{\tilde F}^{\mu\nu}.
                      \label{eq:Y-WT-identity}
\end{equation}
However, the sphaleron process is enhanced in the symmetric phase
and in a restricted region in the broken phase where the expectation 
value of the Higgs is so small that the sphaleron rate is large.
In Fig.~\ref{fig:3}, such a region is that where $v<v_{co}$.
Hence the estimate of the generated baryon number suffers from a
suppression factor $(v_{co}/T)^2$, which could be as small as
$O(10^{-6})$.\par
Later Cohen {\rm et al.} found that the effects of diffusion can 
enhance the baryon asymmetry\cite{CKN-diff}.\  
The diffusion may carry the charge asymmetry into deep symmetric 
phase region. They solved the diffusion equations for various charges,
assuming the profile
\begin{equation}
 \expecv{\phi(z)}=v{{1-\tanh(az)}\over2}\exp\left[
                    -i\Delta\theta{{1-\tanh(az)}\over2}\right]
                          \label{eq:CKN-profile}
\end{equation}
with $\Delta\theta=\pi/2$,
and found that the generated baryon asymmetry becomes enhanced by a
factor of $O(1/\alpha_W^4)\sim10^6$ over the previous estimates, and that
the baryon asymmetry is insensitive to the details of the bubble 
dynamics, such as the wall width $1/a$ and velocity $v_w$.
Once diffusion is taken into account, generation of the baryon 
number is similar to that in the charge transport mechanism.
In this sense, these scenarios are also called 
`nonlocal baryogenesis'. For recent and more elaborated studies on these
subjects, which include the effects of diffusion, 
see Ref.\cite{Joyce}.\par
As we saw in this section, to obtain enough BAU, some extensions of
the MSM would be needed, in both mechanisms.
How these mechanisms contribute to generate the BAU would be 
determined by the dynamics of the EWPT.
We present some results based on the charge transport mechanism.
For thicker bubble walls (smaller $a/T$), the other mechanism
will come into effect, so that the flux for smaller $a/T$ may not
damp so rapidly in practice.
In the illustrative calculations above, the profiles of the $CP$ 
violating bubble wall, (\ref{eq:m-NKC}) and (\ref{eq:CKN-profile}),
were assumed without any reasoning.
They should be determined by the dynamics.
We need to relate $CP$ violation around the bubble wall to
the parameters in the models of electroweak theory, to estimate
quantitatively the generated baryon asymmetry.
%
%
%
%
%-------------
\section{$CP$-violating bubble wall}
\subsection{The model and the equations of motion}
$C$ {\it and} $CP$ violation are among the requirements to generate baryon 
asymmetry starting from a baryon-symmetric universe.
$C$ is violated in any chiral gauge theory such as the standard model,
while $CP$ is violated only through the KM matrix in the MSM.
As we saw in the previous section, some extension of the Higgs sector
of the MSM would be needed to have a sufficient baryon asymmetry
only by the electroweak baryogenesis.
In practice, the amount that the baryon number is generated depends
on the $CP$ violation of the model under consideration.
As we noted, one of the most attractive features of the electroweak 
baryogenesis is that it relies exclusively on physics which can be
verified by experiments in the near future.
Our aim is to develop a formalism by which the generated baryon number
can be evaluated quantitatively once one selects the model lagrangian for
the electroweak theory.
Here we concentrate on the two-Higgs-doublet models, which contain
the minimal supersymmetric standard model (MSSM) as a special case
and is the simplest model to incorporate the Higgs-sector $CP$ violation.
Assuming the effective potential which gives the first-order EWPT,
we determine the functional form of the $CP$ violating phase in the
Higgs fields. For a review of this model and MSSM,
see Ref.~\cite{HiggsHunter}.\par
The most general renormalizable Higgs potential which is
invariant under the $SU(2)_L\times U(1)_Y$ gauge transformation is
\begin{eqnarray}
 V_0&=&
 m_1^2\Phi_1^\dagger\Phi_1 +  m_2^2\Phi_2^\dagger\Phi_2
 +( m_3^2\Phi_1^\dagger\Phi_2 + {\rm h.c.} )                  \nonumber \\
 &+&
 \half\lambda_1(\Phi_1^\dagger\Phi_1)^2+\half\lambda_2(\Phi_2^\dagger\Phi_2)^2
 +\lambda_3(\Phi_1^\dagger\Phi_1)(\Phi_2^\dagger\Phi_2)
 -\lambda_4(\Phi_1^\dagger\Phi_2)(\Phi_2^\dagger\Phi_1)
                                                            \nonumber\\
 &-&
 \left\{ \half\lambda_5(\Phi_1^\dagger\Phi_2)^2
 +\bigl[\lambda_6(\Phi_1^\dagger\Phi_1)+\lambda_7(\Phi_2^\dagger\Phi_2)\bigr]
  (\Phi_1^\dagger\Phi_2) + {\rm h.c.} \right\},      \label{eq:general-V0}
\end{eqnarray}
where the charge assignment of the scalars are $({\bf 2},1)$ for
$SU(2)_L\times U(1)_Y$.
Here the hermiticity of $V_0$ requires
$m_1^2, m_2^2, \lambda_1, \lambda_2, \lambda_3, \lambda_4\in{\bf R}$ 
while in general $m_3^2, \lambda_5, \lambda_6, \lambda_7\in{\bf C}$, 
three of their phases are independent and yield the explicit $CP$ violation.
The Yukawa interactions are generally
\begin{eqnarray}
 {\cal L}_Y&=&
  {\bar q}_L(f_1^{(d)}\Phi_1+f_2^{(d)}\Phi_2)d_R +
  {\bar q}_L(f_1^{(u)}{\tilde\Phi}_1+f_2^{(u)}{\tilde\Phi}_2)u_R
                                       \nonumber\\
 & &
 +{\bar l}_L(f_1^{(e)}\Phi_1+f_2^{(e)}\Phi_2)e_R + \mbox{h.c.},
                       \label{eq:general-Yukawa}
\end{eqnarray}
where ${\tilde\Phi_i}=i\sigma_2\Phi_i^*$ is the charge-conjugated 
Higgs fields and $f_{1,2}^{(u,d,e)}$ are the Yukawa couplings which are
matrices with the generation indices.
If we impose the discrete symmetry,
\begin{eqnarray}
 & &\Phi_1\mapsto \Phi_1,\qquad \Phi_2\mapsto-\Phi_2,\nonumber\\
 & & u_R\mapsto -u_R, \qquad d_R\mapsto d_R,\qquad e_R\mapsto e_R
         \label{eq:GW-dis-sym1}
\end{eqnarray}
or
\begin{eqnarray}
 & &\Phi_1\mapsto \Phi_1,\qquad \Phi_2\mapsto-\Phi_2,\nonumber\\
 & & u_R\mapsto u_R, \qquad d_R\mapsto d_R,\qquad e_R\mapsto e_R
         \label{eq:GW-dis-sym2}
\end{eqnarray}
which forbids the tree-level Higgs-mediated flavor-changing neutral
current (FCNC) interactions\cite{GlashowWeinberg}.\  
the Yukawa interactions in (\ref {eq:general-Yukawa}) are restricted
to the form of
\begin{equation}
 {\cal L}_Y={\bar q}_L f^{(d)}\Phi_1 d_R + 
            {\bar q}_L f^{(u)}{\tilde\Phi}_2 u_R +
            {\bar l}_L f^{(e)}\Phi_1 e_R + \mbox{h.c.}
                      \label{eq:2D-Yukawa1}
\end{equation}
or
\begin{equation}
 {\cal L}_Y={\bar q}_L f^{(d)}\Phi_1 d_R + 
            {\bar q}_L f^{(u)}{\tilde\Phi}_1 u_R +
            {\bar l}_L f^{(e)}\Phi_1 e_R + \mbox{h.c.},
                      \label{eq:2D-Yukawa2}
\end{equation}
and at the same time, in the Higgs potential, it is required
$\lambda_6=\lambda_7=m_3^2=0$. One can introduce $m_3^2\not=0$, which
breaks the discrete symmetry softly, since it does not induce
divergent counterterms of the $\lambda_{6,7}$-type so that there is 
still no tree-level FCNC interactions.\par
In the MSSM, the tree-level Higgs potential is given by
(\ref{eq:general-V0}) with
\begin{eqnarray}
 & &\lambda_1=\lambda_2={1\over 4}(g^2+{g^\prime}^2),\qquad
    \lambda_3={1\over4}(g^2-{g^\prime}^2),\qquad
    \lambda_4={1\over2}g^2,   \nonumber\\
 & &\lambda_5=\lambda_6=\lambda_7=0,   \label{eq:MSSM-V0}
\end{eqnarray}
and the Yukawa coupling is of the type (\ref{eq:2D-Yukawa1}),
where $\Phi_d$ is identified with $\Phi_1$ and $\Phi_u$ with
${\tilde \Phi}_2$.
The $m_3^2$-term in the MSSM comes from the soft supersymmetry
breaking term and is generally complex.\par
When $m_3^2, \lambda_5, \lambda_6, \lambda_7\in{\bf R}$, which
is the case for the MSSM, $CP$ is not violated by the potential.
However, $CP$ can be spontaneously broken if
\begin{equation}
 \lambda_5<0,\qquad
 \absv{{{\lambda_6v_1^2+\lambda_7v_2^2-2m_3^2}\over{2\lambda_5v_1v_2}}}
 <1,                            \label{eq:condition-SCP0}
\end{equation}
where we parameterize the vacuum expectation value of the Higgs as
\begin{equation}
 \expecv{\Phi_1}=
 \pmatrix{0 \cr {1\over{\sqrt{2}}}v_i{ e}^{i\theta_i}\cr},\qquad
 \theta = \theta_1-\theta_2.        \label{eq:def-VEV0}
\end{equation}
This is because, upon inserting (\ref{eq:def-VEV0}) into
(\ref{eq:general-V0}), the $\theta$-dependent part of $V_0$ becomes
\begin{equation}
 V_0= -{{\lambda_5}\over2}v_1^2v_2^2\left[
       \cos\theta+
       {{\lambda_6v_1^2+\lambda_7v_2^2-2m_3^2}\over{2\lambda_5v_1v_2}}
             \right] + \cdots,           \label{eq:th-in-V0}
\end{equation}
which takes the minimum at $\theta\not=n\pi\quad(n\in{\bf Z})$
if the condition (\ref{eq:condition-SCP0}) is satisfied.\footnote{As 
for a solution to the domain wall problem associated with
spontaneous $CP$ violation, see Ref.~\cite{KraussRey}.}
Although the MSSM does not admit the tree-level spontaneous 
$CP$-violation, the radiative corrections can induce the four-point
couplings in such a way that (\ref{eq:condition-SCP0}) is 
fulfilled\cite{Maekawa}.\  When some discrete symmetry is broken 
radiatively, one inevitably gets a pseudo-Goldstone boson, which
is usually light because it acquires its mass only through 
radiative corrections\cite{GeorgiPais}.\   
In the case of spontaneous $CP$ violation in the MSSM, the light scalar 
mass is about $6\mbox{GeV}$ at most, which is phenomenologically
forbidden\cite{Maekawa}.\par
We are interested in $CP$ violation at finite temperature, especially
near the bubble wall created at the EWPT. 
The effective potential of the model at finite temperature, which
depends on the expectation values of all the Higgs fields, would
determine to what extent $CP$ is violated at $T\simeq T_C$ in the broken
phase region.
In fact, the one-loop calculation in the MSSM shows that
$CP$ could be spontaneously violated at high temperature while unbroken
at zero temperature, avoiding the light pseudo-Goldstone 
boson\cite{Comellia,Comellib,Espinosa}.\  
We expect that if we know the global structure of the effective 
potential, we could derive from it the profile of the bubble wall
including $CP$ violation. Here the effective potential, which offers
the first-order EWPT, should reproduce not only $T_C$ and $\varphi_C$
but also various properties of the EWPT such as the latent heat and
surface tension, which are consistent with lattice results.
Since we have no available information about the EWPT in 
the two-Higgs-doublet model, we assume some general form of the 
effective potential at $T_C$. We solve the classical equations of motion 
for the gauge-Higgs system of our model, with the potential replaced
by the postulated effective potential, to determine the possible 
profiles of $CP$ violating bubble wall.
The equations of motion are derived from the lagrangian
\begin{equation}
 {\cal L}=-{1\over4}F^a_{\mu\nu}F^{a\,\mu\nu}-{1\over4}B_{\mu\nu}B^{\mu\nu}
          +\sum_{i=1,2}\left(D_\mu\Phi_i\right)^\dagger D^\mu\Phi_i
          -V_{\rm eff}(\Phi_1,\Phi_2;T),     \label{eq:lagrangian}
\end{equation}
where
$$
  D_\mu\Phi_i(x) \equiv (\del_\mu-ig{{\tau^a}\over2}A^a_\mu(x)
                       -i{{g^\prime}\over2}B_\mu(x))\Phi_i(x).
$$
As long as the EWPT proceeds so that the bubble wall moves keeping
the shape of the critical bubble with a constant velocity,
the static solution connecting the two phases will describe the
bubble wall profile.
When the bubble is spherically symmetric or is macroscopic so that
it is regarded as a planar object, the system is reduced to an 
effective one-dimensional one. Here we assume that the gauge fields
of the solution are pure-gauge type
\begin{equation}
  ig{{\tau^a}\over2}A^a_\mu(x)=\del_\mu U_2(x)U_2^{-1}(x),\qquad
  i{{g^\prime}\over2}B_\mu(x) =\del_\mu U_1(x)U_1^{-1}(x),
          \label{eq:pure-gauge}
\end{equation}
where $U_2$ and $U_1$ are elements of $SU(2)_L$ and $U(1)_Y$, respectively.
This assumption would be justified because 1+1-dimensional gauge 
theories contain no dynamical degrees of freedom of gauge fields.
Further if we find a solution based on this assumption, it will have
the lowest energy, since it has no contribution from the gauge sector
to the energy.
Then we gauge away all the gauge fields, or, in other words, we
fix the gauge in such a way all gauge fields vanish.
Assuming that $U(1)_{em}$ is not broken anywhere, the classical Higgs
scalars are parameterized as
\begin{equation}
  \Phi_i(x)
 =\left(\begin{array}{c} 0 \\
          {1\over{\sqrt2}}\rho_i(x){ e}^{i\theta_i(x)}
        \end{array}\right). \qquad(i=1,2)   \label{eq:classical-Higgs}
\end{equation}
The static one-dimensional equations of motion are
\begin{eqnarray}
 {{d^2\rho_i(z)}\over{dz^2}}-\rho_i(z)\left({{d\theta_i(z)}\over{dz}}\right)^2
  -{{\del V_{\rm eff}}\over{\del\rho_i}} &=& 0,
	\label{eq:rho-z}  \\
 {d\over{dz}}\left(\rho_i^2(z){{d\theta_i(z)}\over{dz}}\right)
  -{{\del V_{\rm eff}}\over{\del\theta_i}} &=& 0,
	\label{eq:theta-z}  \\
  \rho_1^2(z){{d\theta_1(z)}\over{dz}}+\rho_2^2(z){{d\theta_2(z)}\over{dz}}
  &=& 0,
	\label{eq:soureless-z}
\end{eqnarray}
where $z$ is the coordinate perpendicular to the planar bubble wall
and the last equation is the consistency condition for the pure-gauge
ansatz, which may be viewed as the gauge-fixing condition.
We introduce a dimensionless and finite-range variable by 
\begin{equation}
 y = {1\over2}\left(1-\tanh(az)\right),    \label{eq:z-y}
\end{equation}
where $a$ has a dimension of mass and its inverse characterizes the width
of the wall. In terms of this variable, the equations of motion are written 
as
\begin{eqnarray}
 & &
 4a^2y(1-y){d\over{dy}}\left[y(1-y){{d\rho_i(y)}\over{dy}}\right]\nonumber\\
 & &\qquad\qquad
 -4a^2y^2(1-y)^2\rho_i(y)\left({{d\theta_i(y)}\over{dy}}\right)^2
 -{{\del V_{\rm eff}}\over{\del\rho_i}} = 0,     \label{eq:rho-y}  \\
 & &
 4a^2y(1-y){d\over{dy}}\left[y(1-y)\rho_i^2(y){{d\theta_i(y)}\over{dy}}\right]
 -{{\del V_{\rm eff}}\over{\del\theta_i}} = 0,   \label{eq:theta-y}  \\
 & &
  \rho_1^2(y){{d\theta_1(y)}\over{dy}}+\rho_2^2(y){{d\theta_2(y)}\over{dy}}
  = 0.                                       \label{eq:sourceless-y}
\end{eqnarray}
In order to solve these equations, one must know the explicit form of
$V_{\rm eff}$. Because of the gauge invariance, $V_{\rm eff}$ is a function of
$\theta_1-\theta_2$. From this fact, (\ref{eq:theta-y}) with $i=2$ is
automatically satisfied as long as $\rho_i$ and $\theta_i$ satisfy
(\ref{eq:theta-y}) with $i=1$ and (\ref{eq:sourceless-y}).
We determine the effective potential by postulating 
that it is a gauge-invariant polynomial of $\rho_1$, $\rho_2\cos\theta$
and $\rho_2\sin\theta$ up to the fourth order, and that it has two 
degenerate minima, each corresponding to the symmetric and broken phase.
Further, for simplicity, we assume that the modulus of the Higgs field,
$\rho_i(z)$, has a kink shape and that $\rho_1$ and $\rho_2$ have
the same order of width. This assumption may amount to that of the EWPT
proceeding smoothly and accompanying the two Higgs fields which take
nonzero values at about the same temperature.
We expect that the effective potential of the model
is equipped with these features if the EWPT it predicts is first order.\par
For the time being, we consider the case in which $CP$ is not explicitly
violated in $V_{\rm eff}$, so that all parameters are real and $V_{\rm eff}$
depends
on $\theta$ only through $\cos\theta$. We require that (\ref{eq:rho-y}) 
has the kink-type solutions in the absence of $CP$ violation
\begin{equation}
  \rho_i(y) = v_i ( 1 - y ),    \label{eq:rho-kink}
\end{equation}
where
$$
  v_1 = v\cosb,\qquad v_2 = v\sinb.
$$
Then the effective potential takes the form\cite{FKOTTc}
\begin{eqnarray}
& & V_{\rm eff}(\rho_1,\rho_2,\theta)   \nonumber \\
&=&
 (2a^2-\half m_3^2\tan\beta)\rho_1^2+(2a^2-\half m_3^2\cot\beta)\rho_2^2
 + m_3^2\rho_1\rho_2\cos\theta           \nonumber \\
 &-&
  \left\{ A\rho_1^3
 +\left[-2A\cot\beta+D\tan^2\beta+{{4a^2}\over{v\sinb}}(3-{1\over{\cos^2\beta}})
       \right]\rho_1^2\rho_2(\cos\theta) \right.    \nonumber \\
 & &
  \left.
 +\left[A\cot^2\beta-2D\tan\beta+{{4a^2}\over{v\cosb}}(3-{1\over{\sin^2\beta}})
       \right]\rho_1\rho_2^2(\cos\theta) + D\rho_2^3 \right\}\nonumber \\
 &+&
  {{\lambda_1}\over8}\rho_1^4+{{\lambda_2}\over8}\rho_2^4
 +{{\lambda_3-\lambda_4}\over4}\rho_1^2\rho_2^2
 -{{\lambda_5}\over4}\rho_1^2\rho_2^2\cos(2\theta)  \label{eq:full-Veff}\\
 &-&
  {1\over8}\left\{
  \left[{3\over2}\lambda_1\cot\beta-{{\lambda_2}\over2}\tan^3\beta
 +\tilde\lambda_3\tan\beta
 -{{8a^2}\over{v^2\sinb\cosb}}(4-{1\over{\cos^2\beta}})\right]\rho_1^3\rho_2
   \right.    \nonumber \\
 & &
  \left.
 +\left[-{{\lambda_1}\over2}\cot^3\beta + {3\over2}\lambda_2\tan\beta
 +\tilde\lambda_3\cot\beta
 -{{8a^2}\over{v^2\sinb\cosb}}(4-{1\over{\sin^2\beta}})\right]\rho_1\rho_2^3
 \right\}\cos\theta.         \nonumber
\end{eqnarray}
Here $\cos\theta$ in the $\rho^3$-terms is optional.
All the parameters should be regarded as those including radiative and
finite-temperature corrections. Hence even for the MSSM, these 
parameters could be induced in the presence of the soft supersymmetry
breaking terms.
With this potential, the equation for $\theta$ derived from
(\ref{eq:theta-y}) in the kink-background is
\begin{eqnarray}
 & & 
 y^2(1-y)^2{{d^2\theta(y)}\over{dy^2}}+y(1-y)(1-4y){{d\theta(y)}\over{dy}}
          \nonumber\\
 &=&
 [b+c(1-y)^2-e(1-y)]\sin\theta(y) + {d\over2}(1-y)^2\sin(2\theta(y)),
	\label{eq:eq-theta}
\end{eqnarray}
where the parameters are defined by
\begin{eqnarray}
  b&\equiv& -{{m_3^2}\over{4a^2\sinb\cosb}},   \nonumber\\
  c&\equiv& {{v^2}\over{32a^2}}(\lambda_1\cot^2\beta+\lambda_2\tan^2\beta
            +2\tilde\lambda_3) - {1\over{2\sin^2\beta\cos^2\beta}}
      \nonumber \\
   &=&  {{v^2}\over{8a^2}}(\lambda_6\cot\beta + \lambda_7\tan\beta), 
                  \nonumber\\
  d&\equiv& {{\lambda_5 v^2}\over{4a^2}}.   \nonumber   \\
  e&\equiv& {v\over{4a^2\sin^2\beta\cos^2\beta}}
             \left( A\cos^3\beta+D\sin^3\beta-{{4a^2}\over v} \right)
                   \nonumber\\
   &=& -{v\over{4a^2}}\left({B\over\sinb}+{C\over\cosb}\right).
                       \label{eq:def-bcd}
\end{eqnarray}
We refer to the solution of this equation as that with the `kink ansatz'.
When $\theta(y)$ becomes of $O(1)$, actual solution will no longer
satisfy the kink ansatz for $\rho(y)$, (\ref{eq:rho-kink}).
Now we show some solutions found numerically for various boundary 
conditions\cite{FKOTTc,FKOTd}.\  
Similar attempts were made by Cline {\rm et al.}\cite{Cline}
\subsection{Solutions with spontaneous $CP$ violation}
The possible boundary conditions on $\theta(y)$ depend on the 
parameters $b,c,d$ and $e$. 
Since we concentrate on the potential without explicit $CP$ violation,
the boundary condition in the broken phase ($y=0$) is either
spontaneously generated $\theta_0\equiv\theta(y=0)\not=0$ or
$\theta=n\pi$. The former case is realized when the parameters satisfy
\begin{equation}
  d<0 \quad\mbox{and}\quad \absv{b+c-e}<-d, \label{eq:cond-SCPtheta}
\end{equation}
which corresponds to (\ref{eq:condition-SCP0}).
Then the boundary value $\theta_0$ is given by
\begin{equation}
 \cos\theta_0 = - {{b+c-e}\over d}.   \label{eq:theta0}
\end{equation}
Although one might consider that $\theta_1\equiv\theta(y=1)$ can take
any value in the symmetric phase since the Higgs scalars vanish there,
the finiteness of the energy density of the solution requires that
$\theta_1=n\pi$.\par
We found several solutions, first assuming the power series solution,
for these two types of boundary conditions.
This assumption, which is not essential, somehow restricts the 
parameters in the potential. We present some of the numerical solutions.
\par\noindent
(i) {\it solution violating $CP$ spontaneously in the broken phase}\par
\begin{figure}
 \epsfxsize = 7cm
 \centerline{\epsfbox{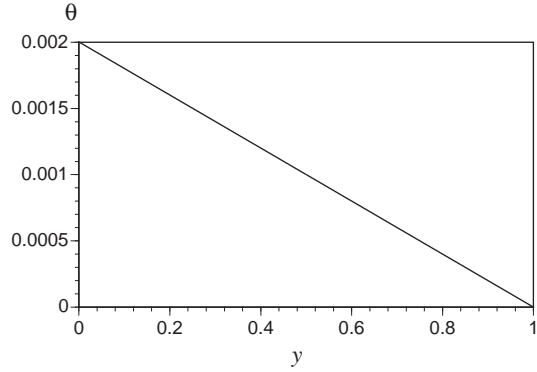}}
 \caption{The numerical solution of $\theta(y)$ for $\theta(0)=0.002$\
 and $\theta(1)=0$. The parameters are $(b,c,e)=(3,7,7)$.}
 \label{fig:9}
\end{figure}

\begin{figure}
 \epsfxsize = 7cm
 \centerline{\epsfbox{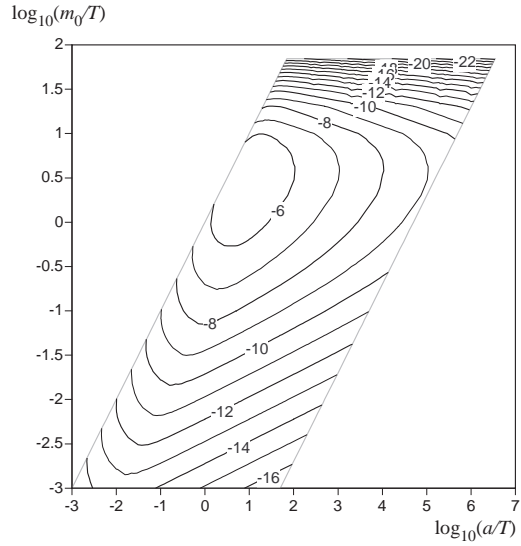}}
 \caption{Contour plot of the chiral charge flux, normalized as
 $\log_{10}\left[-F_Q/(v_w T^3(Q_L-Q_R))\right]$ for the profile shown
 in Fig.~9. % Fig.~\ref{fig:9}. 
 Here we take $v_w=0.58$ and $T=100\mbox{GeV}$.}
 \label{fig:10}
\end{figure}

\noindent
When the condition (\ref{eq:cond-SCPtheta}) is satisfied, 
there could be a solution with the boundary condition 
$\theta_0\not=n\pi$ and $\theta_1=m\pi$.
We found several solutions and two of them are depicted in 
Ref.~\cite{FKOTTc}.\  
Both of them satisfy $\theta_1=0$, while one of them
has small $\theta_0\not=0$, which may be consistent with the present 
experimental bound.
The other has $\theta_0=1$. Such a solution may be realized when the 
finite-temperature effects enhance spontaneously generated $CP$ asymmetry,
which is restored at zero temperature as shown in the 
MSSM\cite{Comellia}.\  
But the parameter space admitting such a possibility seems rather
restricted\cite{Espinosa},\  so that the former case with small 
$\theta_0$ may be more likely to occur.
We show the profile for $(b,c,e)=(3,7,7)$, $\theta_0=0.002$ and $d$ 
is determined by (\ref{eq:theta0}) in Fig.~\ref{fig:9} and the chiral 
charge flux for the profile at $v_w=0.58$ and $T=100\mbox{GeV}$
in Fig.~\ref{fig:10}\  calculated by the numerical method of 
Ref.~\cite{FKOTb}.\  
This is almost a straight line, because for small $\theta_0$, 
$\theta(y)=\theta_0(1-y)$ is an approximate
solution to the linearized version of (\ref{eq:eq-theta}).
%
%%%%%%% Fig.9 %%%%%%%%
%
%%%%%%% Fig.10 %%%%%%%%
As shown in Fig.~\ref{fig:10}, the magnitude of the chiral charge flux
is reduced to about $O(10^{-3})$ of that in Fig.~\ref{fig:7}, which
corresponds to the postulated profile (\ref{eq:m-NKC}) with 
maximal $CP$ violation.
Hence, in this case, even if the forward-scattering is enhanced
to give an extra factor of $O(10^3)$ so that the baryon asymmetry is 
the optimal value (\ref{eq:opt-BAU}), the parameter space is 
rather restricted to generate the present BAU only with the charge
transport mechanism.\par\noindent
(ii) {\it solution conserving $CP$ in the broken phase}\par
\begin{figure}
 \epsfxsize = 7cm
 \centerline{\epsfbox{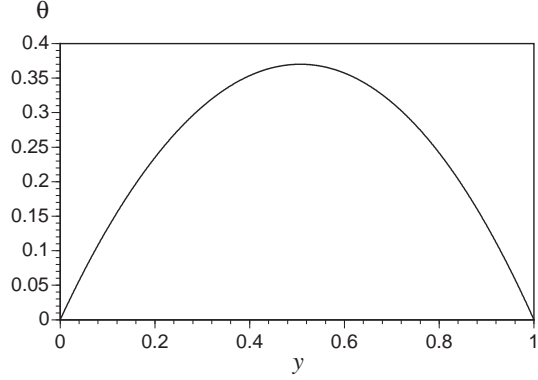}}
 \caption{The numerical solution of $\theta(y)$ for
 $\theta(0)=\theta(1)=0$. The parameters are $(b,c,d,e)=(3,12.2,-2,12.2)$.}
 \label{fig:11}
\end{figure}

\begin{figure}
 \epsfxsize = 7cm
 \centerline{\epsfbox{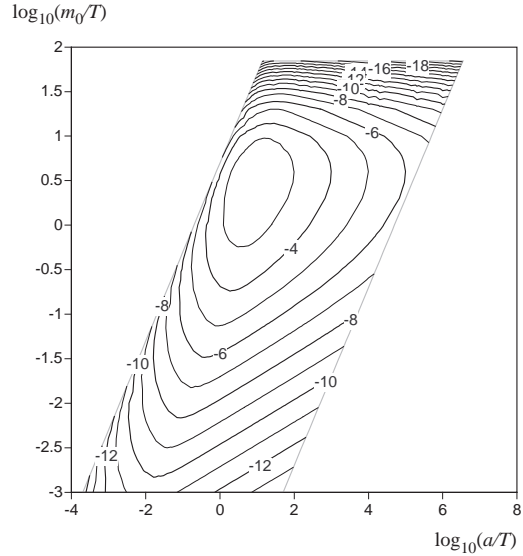}}
 \caption{Contour plot of the chiral charge flux, normalized as
 $\log_{10}\left[-F_Q/(v_w T^3(Q_L-Q_R))\right]$ for the profile shown
 in Fig.~11. % Fig.~\ref{fig:11}. 
 Here we take $v_w=0.58$ and $T=100\mbox{GeV}$.}
 \label{fig:12}
\end{figure}

\noindent 
In this case the boundary condition are $\theta_0=n\pi$ and $\theta_1=m\pi$.
Then there are the trivial solution $\theta(y)=n\pi$ with the 
kink-type $\rho(y)$. Besides these, we found an interesting solution
which violates $CP$ in the intermediate range near the bubble wall.
If the effective potential admits the condition for the spontaneous
$CP$ violation (\ref{eq:condition-SCP0}) to be satisfied for intermediate
$v_i$, as shown in the contour plot of $V_{\rm eff}$ in
Ref.~\cite{FKOTTc}, such a solution exists.
Note that this condition is weaker than (\ref{eq:cond-SCPtheta}),
which is required for spontaneous $CP$ violation in the broken phase region.
Both conditions need $\lambda_5<0$. Unless the model has
tree-level negative $\lambda_5$, it would be difficult to obtain negative
$\lambda_5$ starting from $\lambda_5\ge0$ at the tree-level,
since only the fermions such as the gauginos in the MSSM contribute
negatively to $\lambda_5$, while contributions from the bosons and
fermions at finite temperature are positive.
We present a solution of this type found in Ref.~\cite{FKOTTc} 
and the chiral charge flux for it, in Fig.~\ref{fig:11} and
Fig.~\ref{fig:12}, respectively.
This type of solution may exist for large $e$, which is the coefficient
of $\rho^3\cos\theta$-term in the effective potential.
As we noted earlier, the $\rho^3$-term arises from the boson loops whose
mass at $\varphi=0$ vanishes.
In the MSSM, the squark loop might give such a term if its mass is very
small for $\varphi=0$.
%
%%%%%%% Fig.11 %%%%%%%%
%
%%%%%%% Fig.12 %%%%%%%%
Although the maximum of $\theta(y)$ is about $0.3$, the flux is 
comparable to that in Fig.~\ref{fig:7}, so that this profile
could generate sufficient baryon asymmetry.
Since this solution and the trivial solution $\theta(y)=0$
satisfy the same equation and the same boundary conditions, both would
appear at the EWPT, but with different probabilities.
One can determine the difference in the nucleation rate by comparing
the free energy of the critical bubbles.
The energy density per unit area of a bubble is given by
\begin{eqnarray}
 {\cal E}&=&\int_{-\infty}^\infty dz \left\{
  \half\sum_{i=1,2}\left[\left({{d\rho_i}\over{dz}}\right)^2
    +\rho_i^2 \left({{d\theta_i}\over{dz}}\right)^2 \right]
   + V_{\rm eff}(\rho_1,\rho_2,\theta) \right\}      \nonumber\\
 &=&
  \int_0^1 dy \left\{
  ay(1-y)\sum_{i=1,2}\left[\left({{d\rho_i}\over{dy}}\right)^2
           +\rho_i^2 \left({{d\theta_i}\over{dy}}\right)^2\right]
              \right.\nonumber\\
 & &\qquad\qquad\qquad \left. 
  +{1\over{2ay(1-y)}}V_{\rm eff}(\rho_1,\rho_2,\theta) \right\}.      
           \label{eq:energy-y}
\end{eqnarray}
The difference in the energy density is\cite{FKOTTc}
\begin{equation}
  \Delta{\cal E} = {\cal E} - \left.{\cal E}\right|_{\theta=0}
                 = - 2.056\times 10^{-3}\,av^2\sin^2\beta\cos^2\beta.
           \label{eq:energy-density-diff}
\end{equation}
For the critical bubble of radius $R_C$, the $CP$-violating bubble
will be nucleated with probability larger than the trivial one by
the factor
\begin{equation}
  {\rm exp}\left( -{{4\pi R_C^2\Delta{\cal E}}\over{T_C}} \right).
        \label{eq:prob-CP-viol-bubble}
\end{equation}
According to the estimation in the massless two-Higgs-doublet 
model\cite{FKT},\  the radius of the critical bubble is given by
$\sqrt{3F_C/(4\pi av^2)}$, where $F_C$ is the free energy of the
critical bubble and is found to be about $145T$. Then the exponent in
(\ref{eq:prob-CP-viol-bubble}) is $0.89\sin^2\beta\cos^2\beta$,
irrespective of the wall width.
For $\tan\beta=1$, the $CP$-violating bubbles are created 
about $1.22$ times more than the trivial ones.\par
Now we comment on the baryogenesis in the absence of explicit $CP$
violation. Since the effective potential is an even function of
$\theta$, $-\theta(y)$ is also a solution to the equations of motion
if $\theta(y)$ is. The energy densities of the both bubbles with
$\theta(y)$ and $-\theta(y)$ --- we refer to them as `positive bubble'
and `negative bubble', respectively --- are equal to each other.
Thus their nucleation rates are also the same, so that
the net generated baryon number will be zero on the average.
This degeneracy of the energy density will be resolved once explicit 
$CP$ violation is taken into account.
It was shown that in the MSSM, the explicit $CP$ violation in the soft 
supersymmetry breaking terms induces a $m_3^2$-type term proportional
to $\sin\theta$ in the effective potential\cite{Comellib},\  
If the difference in the free energy of the two bubbles is $\Delta F$,
the ratio of their number will be 
\begin{equation}
 {{N_+}\over{N_-}} = { e}^{-\Delta F/T},  \label{eq:number-ratio}
\end{equation}
and the net baryon number is 
\begin{equation}
 {{n_B}\over s}=\left({{n_B}\over s}\right)_0
                {{N_+-N_-}\over{N_++N_-}}
         \simeq \left({{n_B}\over s}\right)_0{{\Delta F}\over T},      
      \label{eq:net-B-asymmetry}
\end{equation}
where $(n_B/s)_0$ is the baryon asymmetry generated by each bubble.
Comelli {\rm et al.} showed that the explicit $CP$ violation, which does
not yield the neutron electric dipole moment beyond the experimental bound,
splits the free energy as large as $\Delta F/T\simeq 
10^{-2\sim-1}$\cite{Comellib}.\  
If this is the case, the BAU is explained by the electroweak theory
as long as $(n_B/s)_0\gtsim10^{-9}$.
\subsection{Solutions with explicit $CP$ violation}
As we noted, explicit $CP$ violation is necessary to have nonzero
baryon asymmetry. We found that even if it is very small, it 
nonperturbatively yields an energy gap between the positive and negative 
bubbles\cite{FKOTd}.\par
Here we consider only the phase in the $m_3^2$-term of the type
$m_3^2({ e}^{-i\delta}\Phi_1^\dagger\Phi_2+\mbox{h.c.})$, which is
indeed induced in the MSSM. Then the equation for $\theta$ with the kink 
ansatz is now
\begin{eqnarray}
  & &y^2(1-y)^2{{d^2\theta(y)}\over{dy^2}}+y(1-y)(1-4y){{d\theta(y)}\over{dy}}
         \label{eq:eq-theta-ecp}\\
 &=&b\sin(\delta+\theta(y))+\bigl[c(1-y)^2-e(1-y)\bigr]\sin\theta(y)
+ {d\over2}(1-y)^2\sin(2\theta(y)),
	     \nonumber
\end{eqnarray}
where the parameters are defined by (\ref{eq:def-bcd}).
Just as the case with spontaneous $CP$ violation above, the boundary 
conditions are determined by the parameters in the potential.
We found two types of numerical solutions.\par
One type of solution is that of $\theta(y)=O(\delta)$ in the whole region
and $\theta(1)=-\delta$. Note that $-\theta(y)$ is no longer a solution
to (\ref{eq:eq-theta-ecp}). Hence there is no cancellation, but the
generate baryon asymmetry would be the same order as that shown in 
Fig.~\ref{fig:10}, if $\delta=O(10^{-3})$.\par
\begin{figure}
 \epsfxsize = 7cm
 \centerline{\epsfbox{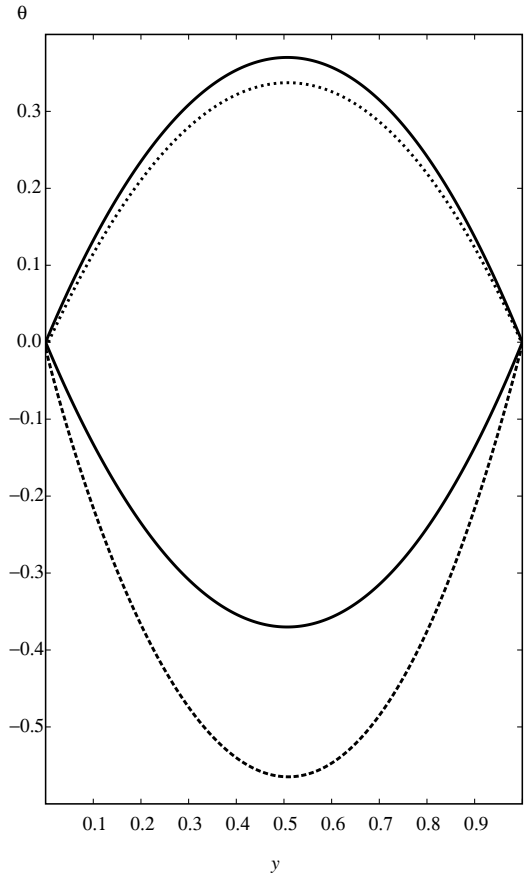}}
 \caption{The numerical solutions $\theta^-(y)$ (the dashed curve) and
 $\theta^-(y)$ (the dotted curve) for $\delta=0.0025$ and 
 $(b,c,d,e)=(2.98005,12.178375,-2,12.2)$, which satisfy the boundary 
 conditions, $\theta_0^{\pm}\simeq-0.311\delta$ and 
 $\theta_1^{\pm}=-\delta$.
 The solid curves are the degenerate solutions for $\delta=0$ and
 $(b,c,d,e)=(3,12.2,-2,12.2)$ with $\theta_0=\theta_1=0$.}
 \label{fig:13}
\end{figure}

The other solution is similar to that in Fig.~\ref{fig:11}, which connects
$\theta(0)\simeq\delta$ and $\theta(1)=-\delta$ and grows to
$\theta(y)\sim-0.6$ at $y\simeq0.5$.
For sufficiently small $\delta$ ($\sim10^{-3}$),  it has a partner
satisfying the same boundary condition but with opposite sign in
the intermediate region.
For larger $\delta$, the partner can no longer exist.
We calculated their energy density and found that
for $\delta=0.0025$ and $(b,c,d,e)=(2.98005,12.178375,-2,12.2)$,
\begin{equation}
  \Delta{\cal E}\equiv {\cal E}[\theta^-]-{\cal E}[\theta^+]
   =- 1.917\times10^{-2}av^2\sin^2\beta\cos^2\beta,
\end{equation}
where $\theta^{-(+)}(y)$ denotes the solution with lower (higher)
energy density.
They are depicted in Fig.~\ref{fig:13}, which shows that 
$\theta^{\pm}(y)$ deviates from the solutions with $\delta=0$ by
$O(0.1)$ near the bubble wall ($y\sim0.5$) in spite of the small $\delta$.
%
%%%%%%% Fig.13 %%%%%%%%
The ratio of the nucleation rates of the bubbles is
\begin{equation}
 {{N_-}\over{N_+}}
 =\exp\left( -{{4\pi R_C^2\Delta{\cal E}}\over{T_C}}\right)=8.05,
                   \label{eq:ratio-ecp}
\end{equation}
where we put $\tan\beta=1$.
Then the net generated baryon asymmetry will be
\begin{equation}
 {{n_B}\over s}=
 \left({{n_B}\over s}\right)_+{{N_+}\over{N_++N_-}}
 +\left({{n_B}\over s}\right)_-{{N_-}\over{N_++N_-}}
 \simeq
 \left({{n_B}\over s}\right)_+{{N_+-N_-}\over{N_++N_-}},
                        \label{eq:net-B-asymmetry-ecp}
\end{equation}
where we used $(n_B/s)_-\simeq-(n_B/s)_+$.\footnote{When the 
discrepancy between $\theta^+(y)$ and $-\theta^-(y)$ is large,
this approximation is invalid.}
Since the chiral charge flux for this profiles are expected to be
the same order as that in Fig.~\ref{fig:12}, we will obtain
sufficient baryon asymmetry.\par
All the solutions presented in this section are based on the kink 
ansatz. In fact, the solutions other than the trivial ones
$\theta(y)=n\pi$ are not exact solutions to the full equations of motion.
We also get solutions without this ansatz, which would have
lower energy. In fact we found some numerical solutions,
which have $\theta(y)$ of $O(1)$ so that $\rho_i(y)$ are no longer
kink shape\cite{FKOTe}.\  
Such solutions may exist for a broader parameter region.
We have concentrated on static solutions.
If the EWPT accompanies exploding bubbles, we would need to solve
the time-dependent equations, and if the viscosity of the medium
cannot be ignored, we would have to solve the equations with 
frictions or random forces and then the wall profile might not give
a monotonical $\rho$ like those obtained here.
These extensions of the equations of motion are to be studied 
in the future.
%
%
%
%
%-------------
\section{Discussion}
In this paper, we briefly reviewed the scenario of the electroweak 
baryogenesis. For it to be successful, some extensions of the MSM would
be needed to have $CP$ violation in the Higgs sector.
As one such model, we analyzed a two-Higgs-doublet model, including
the MSSM, and tried to relate $CP$ violation to the baryon asymmetry
generated at the EWPT.
The main attraction of the electroweak baryogenesis is that
it relies exclusively on physics which can be tested by present or
near future experiments.
In order to bridge the gap between detectable microphysics and
the baryon asymmetry observed in the universe, we still have to know
various aspects of the electroweak theory, in particular the EWPT and
$CP$ violation in the model.\par
In principle, one can predict how much the baryon asymmetry is 
generated at the EWPT, once the renormalized lagrangian of the 
electroweak theory is given. From the lagrangian, one can construct
the effective potential at finite temperature near the phase transition.
One may use higher-order perturbation or nonperturbative methods such
as the lattice simulation. As we saw in \S~4, the studies
have been limited to those of the MSM. We hope these efforts to be
extended to the extended versions of the MSM.
The effective potential gives information about the EWPT, such
as its order, the transition temperature, and the latent heat
and surface tension if it is first order.
If the EWPT is not first order, the baryon number would be washed out
except for the portion proportional to the primordial $B-L$.
Even if the EWPT is first order, the unsuppressed sphaleron process
would erase the baryon asymmetry.
The Higgs mass is bounded from above, requiring that the EWPT be first 
order or that the sphaleron processes decouple after the EWPT.
The upper bound seems inconsistent with the present lower bound of
the Higgs scalar in the MSM. More detailed studies of the EWPT in
the extended models will provide new constraints on the lightest
Higgs boson. When the EWPT is first order, one can describe how
the transition proceeds. If the phase transition accompanies 
nucleation and successive growth of the bubbles of the broken phase
in the symmetric phase, the mechanisms presented in \S~5 will
yield the baryon asymmetry. The estimate of the generated baryon number
depends on the velocity and thickness of the bubble wall and $CP$ 
violation around it, both of which should be dynamically 
determined.\par
When we analyzed the equations of motion for the Higgs fields in 
\S~6, we did not completely fix the parameters in the potential.
In practice, these are fixed once one chooses the model so that
the type of profile is realized would also be determined.
Under the assumptions made there, $\lambda_5<0$ might be necessary to
have large $CP$ violation, while it may be a weaker condition for spontaneous
$CP$ violation in the broken phase. Then a very small explicit $CP$ violation
is needed to have nonzero net baryon asymmetry.
In any case, if more detailed features of the EWPT in the 
two-Higgs-doublet model become available, one could calculate the
generated baryon asymmetry, which in turn constrains some parameters
in the model to explain the present BAU only with the help of the 
electroweak theory.
\section*{Acknowledgements}
I wish to thank K.~Inoue for encouraging me to write this paper.
I am grateful to my collaborators, A.~Kakuto, S.~Otsuki, K.~Takenaga
and F.~Toyoda. I am indebted to them for the works on the subject of
electroweak baryogenesis.
This work was supported in part by the Grant-in-Aid for Scientific
Research from the Ministry of Education, Science and Culture,
No.08740213.
%
%
%-------------
\appendix
\section{Summary of the Big Bang Cosmology}
In this appendix, we summarize the basics of the standard big bang
cosmology and estimate various time scales at temperature
near the electroweak phase transition.
The most general form of homogeneous and isotropic space is the 
Friedmann-Robertson-Walker (FRW) metric, given by
\begin{equation}
 ds^2 = dt^2 - R^2(t)\left[ 
 {{dr^2}\over{1-kr^2}}+r^2(d\theta^2+\sin^2\theta\,d\phi^2)
                     \right],          \label{eq:FRW-metric}
\end{equation}
where $R(t)$ is the scale factor in the comoving coordinate.
$k$ characterizes the topology of the space and $k=1,0,-1$ correspond to
closed, flat and open space, respectively.
The Einstein equation leads to
\begin{eqnarray}
  \left({{\dot R}\over R}\right)^2+{k\over{R^2}}-{\Lambda\over3}&=&
  {{8\pi G_N}\over3}\rho,        \label{eq:FRW1}\\
  {{\ddot{R}}\over R}-{\Lambda\over3}&=&
  -{{4\pi G_N}\over3}(\rho+3p),   \label{eq:FRW2}
\end{eqnarray}
where $\rho$ is the energy density, $p$ is the isotropic pressure, and
$\Lambda$ is the cosmological constant.
$\rho$ and $p$ are related by the equation of state; 
$p=\gamma\rho$, where $\gamma=1/3$ for a radiation-dominated universe 
and $\gamma\llt 1$ for a matter-dominated universe.
Eliminating $\Lambda$ from (\ref{eq:FRW1}) and (\ref{eq:FRW2}), we have
\begin{equation}
 (R^3\rho)^{\cdot}+3R^2{\dot R}p=0,        \label{eq:FRW3}
\end{equation}
from which we obtain by use of the equation of state
\begin{equation}
 \rho R^{3(\gamma+1)}=\mbox{constant}.      \label{eq:conservation}
\end{equation}
Because of
\begin{equation}
  \int{{d^3{\mib k}}\over{(2\pi)^3}}\absv{{\mib k}}
      {1\over{{ e}^{\absv{{\mib k}}/T}\mp 1}}
 =\left\{\begin{array}{l}
          \displaystyle{{{\pi^2}\over{30}}T^4}, \\
          \displaystyle{{7\over8}{{\pi^2}\over{30}}T^4},
         \end{array}     \right.       \label{eq:energy-density}
\end{equation}
the energy density in a radiation-dominated universe is given by
\begin{equation}
 \rho(T)={{\pi^2}\over{30}}g_*T^4,     \label{eq:rho-RD}
\end{equation}
where $g_*$ represents the effective degrees of freedom at temperature $T$,
\begin{equation}
 g_* \equiv \sum_B g_B + {7\over8}\sum_F g_F.  \label{eq:def-g}
\end{equation}
Here $g_{B(F)}$ counts the spin and internal degrees of freedom of 
the bosons (fermions).
For the standard model with $N_f$ generations and $m$ Higgs doublets,
\begin{equation}
  g_*=24+4m+{7\over8}\times30N_f,         \label{eq:g-for-SM}
\end{equation}
so that $g_*=106.75$ for the MSM.
In a radiation-dominated universe, the Hubble parameter is 
approximately given, from (\ref{eq:FRW1}) with $\Lambda=0$, by
\begin{equation}
  H \simeq\sqrt{{{8\pi G_N}\over3}\rho}\simeq
          1.66\sqrt{g_*}{{T^2}\over{m_{Pl}}},       \label{eq:hubble}
\end{equation}
where $m_{Pl}=\sqrt{G_N}=1.22\times10^{19}\mbox{GeV}$ is the Planck
mass.\par
The entropy density at present is related to the photon density as 
follows.
As long as the local equilibrium is maintained, the law of 
thermodynamics implies
\begin{equation}
 d(\rho_i V) = T_idS_i - p_idV + \mu_idN_i,  \label{eq:thermodyn}
\end{equation}
where the index $i$ denotes the particle species and $S_i$ is the 
entropy in a comoving volume $V$.
This equation, together with the Gibbs-Duhem relation
\begin{equation}
 S_idT_i - Vdp_i + N_id\mu_i = 0,        \label{eq:GibbsDuhem}
\end{equation}
lead to, up to a constant,
\begin{equation}
 s_i\equiv {{S_i}\over{T_i}} =
 {{\rho_i+p_i+\mu_iN_i/V}\over{T_i}}.           \label{eq:def-si}
\end{equation}
Neglecting the chemical potentials ($\mu_i\llt T_i$), the
entropy density is given by
\begin{equation}
 s = \sum_i s_i = \sum_i{{\rho_i+p_i}\over T_i}.
                \label{eq:def-s}
\end{equation}
Since the entropy is dominated by relativistic particles,
\begin{equation}
 s = {{2\pi^2}\over{45}}g_{*S}T^3,    \label{eq:s-T}
\end{equation}
where
\begin{equation}
 g_{*S}\equiv \sum_B g_B\left({{T_B}\over T}\right)^3 +
              {7\over8}\sum_F g_F\left({{T_F}\over T}\right)^3.
              \label{eq:def-gS}
\end{equation}
We can ignore the difference between $T_i$ and $T$ at high
temperatures as shown below, while the difference is significant for
$T<1\mbox{MeV}$.
At such low temperatures, $\nu$-$\bar\nu$ annihilation decouples and 
the entropy in ${\rm e}^\pm$ pairs is transferred only to the photons.
The entropy conservation implies
\begin{equation}
 \left({{T_\nu}\over{T_\gamma}}\right)^3 =
 g_\gamma/(g_\gamma+{7\over8}g_e) = {4\over{11}}.
             \label{eq:T-ratio}
\end{equation}
Hence today we have
\begin{equation}
  g_{*S}=2 + {7\over8}\times{4\over{11}}\times2\times3 = 3.91,
             \label{eq:gS-today}
\end{equation}
which yields
\begin{equation}
  s = 7.04\cdot n_\gamma         \label{eq:s-photon}
\end{equation}
with $n_\gamma$ being given by (\ref{eq:particle-density}).\par
Now we shall estimate time scales of various interactions near
$T_C\simeq100\mbox{GeV}$ of the EWPT.
Given a cross section $\sigma$ of some interaction, the mean free path
$\lambda$ is estimated as
\begin{equation}
  \lambda\cdot\sigma = {1\over n},     \label{eq:def-mfp}
\end{equation}
where $n$ is the density of the particles which participate the 
interaction.
At temperature $T$, $n$ of a massless particle whose degree of 
freedom is $g$ is
\begin{equation}
 n=g\int{{d^3{\mib k}}\over{(2\pi)^3}}
      {1\over{{ e}^{\absv{{\mib k}}/T}\mp 1}}
  =\left\{\begin{array}{l}
           \displaystyle{{{\zeta(3)}\over{\pi^2}}gT^3}, \\
           \displaystyle{{3\over4}{{\zeta(3)}\over{\pi^2}}gT^3},
          \end{array}     \right.     \label{eq:particle-density}
\end{equation}
where $\zeta(3)=1.2020569\cdots$.
The mean free time $\bar t$ of a particle of mass $m$ and
energy $E$ is given by
\begin{equation}
 {\bar t} = {\lambda\over v}={\lambda\over{\sqrt{1-m^2/E^2}}}.
                                       \label{eq:def-mft}
\end{equation}
For $E\ggt m$, ${\bar t}\simeq\lambda$.
Since the cross section of the interaction with the fine structure 
constant $\alpha$ at the center-of-mass energy $\sqrt{s}$ is
$\sigma\simeq \alpha^2/s$, the mean free path at $T$ is
\begin{equation}
 \lambda\simeq {{10}\over{g_*T^3}}\cdot{{T^2}\over{\alpha^2}}
        \simeq {1\over{10\alpha^2 T}},     \label{eq:approx-mfp}
\end{equation}
where we have used $g_*\simeq100$ and $s\sim T^2$.\par
If we take $T=100\mbox{GeV}$,
\begin{equation}
 \begin{array}{ll}
   \displaystyle{\lambda_s \simeq
                 {1\over{10^3\alpha_s}}\sim 0.1\mbox{GeV}^{-1}}
     &\mbox{for strong interactions},\\
   \displaystyle{\lambda_{EW} \simeq
                 {1\over{10^3\alpha_W}}\sim 1\mbox{GeV}^{-1}}
     &\mbox{for electroweak interactions},\\
   \displaystyle{\lambda_Y \simeq
                 \left({{m_W}\over{m_f}}\right)^4\lambda_{EW}}
     &\mbox{for Yukawa interactions},
 \end{array}          \label{eq:elementary-mfp}
\end{equation}
where we have used $\alpha_s(m_Z)=0.117\pm0.005$ and 
$\alpha_W=\alpha_{QED}/\sin^2\theta_W\simeq 1/30$.
At this temperature, the time scale of the universe expansion is,
from (\ref{eq:hubble}),
\begin{equation}
 H^{-1}(T)\simeq 10^{14}\mbox{GeV}^{-1}.  \label{eq:HatEWPT}
\end{equation}
The time scale of the sphaleron process is
\begin{equation}
 {\bar t}_{\rm sph}\simeq (\Gamma_{\rm sph}/n)^{-1}\sim 10^5\mbox{GeV}^{-1}.
                   \label{eq:sph-scale}
\end{equation}
The thickness and velocity of the bubble nucleated at the EWPT
are
\begin{equation}
  l_w\simeq{{1\sim40}\over T}\simeq 0.01\sim 0.4 \mbox{GeV}^{-1}
                \label{eq:wall-thickness}
\end{equation}
and
\begin{equation}
  v_w\simeq 0.1\sim 0.9,           \label{eq:wall-velocity}
\end{equation}
respectively\cite{Liu}.\  
Although $l_w$ and $v_w$ are correlated, we adopt as the scale of the 
EWPT,
\begin{equation}
  t_{\rm wall}={{l_w}\over{v_w}}\simeq 0.01\sim 4 \mbox{GeV}^{-1}.
                                   \label{eq:t-w}
\end{equation}
From these we see the following:
\begin{enumerate}
 \item  All the particles are in {\it kinetic equilibrium} at the same 
 temperature, because of $H^{-1}\ggt {\bar t}_{EW}$, far from the 
 bubble wall.
 \item  The Yukawa interactions of the light fermions 
 ($m_f<0.1\mbox{GeV}$) are out of {\it chemical equilibrium}.
 \item  Some of the flavor-changing interactions are out of chemical 
 equilibrium because of small KM matrix elements.
 \item  Since for the leptons $\lambda_Y>\lambda_{EW}\ggt l_w$, 
 the leptons propagate almost freely before and after the scattering off
 the bubble wall.
 \item  Because of $t_{\rm wall}\llt {\bar t}_{\rm sph}$, the sphaleron 
 process is out of chemical equilibrium near the bubble wall.
\end{enumerate}
\section{Derivation of the Baryon-Number-Changing Rate}
Suppose that there are states with discrete labels, $i$.
Let $P(i;t)$ be the probability to find the system in the state $i$
at time $t$, and $\Gamma_{i\rightarrow j}$ be the transition 
probability from state $i$ to $j$ per unit time.
Then the following master equation holds:
\begin{equation}
 P(i;t+\Delta t) =
 -\sum_{j\not=i}P(i;t)\Gamma_{i\rightarrow j}\Delta t + 
  \sum_{j\not=i}P(j;t)\Gamma_{j\rightarrow i}\Delta t + P(i;t).
                \label{eq:master-eq}
\end{equation}
For a steady state, detailed balance is maintained, that is,
$P(i;t)$ is independent of $t$; $P(i;t)=P_{\rm eq}(i)$ for any $t$.
Then (\ref{eq:master-eq}) is reduced to the detailed balance equation;
\begin{equation}
  \sum_{j\not=i}P_{\rm eq}(i)\Gamma_{i\rightarrow j}
 =\sum_{j\not=i}P_{\rm eq}(j)\Gamma_{j\rightarrow i}.   \label{eq:DB0}
\end{equation}
This implies for the case of baryon number changing processes
\begin{eqnarray}
 & &
 \sum_{n=1}^\infty P_{\rm eq}(B)
  \left(\Gamma_{B\rightarrow B+n}+\Gamma_{B\rightarrow B-n}\right)
    \nonumber\\
 &=&
\sum_{n=1}^\infty \left[ P_{\rm eq}(B+n)\Gamma_{B+n\rightarrow B} +
                          P_{\rm eq}(B-n)\Gamma_{B-n\rightarrow B} \right],
                                                 \label{eq:DB}
\end{eqnarray}
where $P_{\rm eq}(B)\propto{ e}^{-F_B/T}$ with $F_B$ being the free energy
of the state with baryon number $B$.
The transition rates with $\Delta B=\pm1$ are given by 
$\Gamma_+=\Gamma_{B\rightarrow B+1}$
and $\Gamma_-=\Gamma_{B\rightarrow B-1}$ and the others are 
approximated by $\Gamma_{B\rightarrow B+n}\simeq\Gamma_+^n$ and
$\Gamma_{B\rightarrow B-n}\simeq\Gamma_-^n$. 
Since $F_{B+n}=F_B+n\mu_B$,  (\ref{eq:DB}) yields
\begin{equation}
 \sum_{n=1}^\infty \left[\Gamma_+^n + \Gamma_-^n\right] \simeq
 \sum_{n=1}^\infty \left[({ e}^{-\mu_B/T}\Gamma_-)^n +
                         ({ e}^{\mu_B/T}\Gamma_+)^n\right].
                               \label{eq:DB2}
\end{equation}
If $\Gamma_\pm\llt 1$, only the contribution from $n=1$ in the sum
is dominant so that
$\Gamma_+ +\Gamma_-\simeq
 { e}^{-\mu_B/T}\Gamma_-+{ e}^{\mu_B/T}\Gamma_+$, which leads to
\begin{equation}
  {{\Gamma_+}\over{\Gamma_-}}\simeq { e}^{-\mu_B/T}. 
                            \label{eq:rasio-Gamma}
\end{equation}
By definition, ${\dot n}_B=\Gamma_+-\Gamma_-$ if we use $\Gamma_\pm$
as the rate per unit volume and unit time. That is approximately 
given by the sphaleron rate, $\Gamma_+\sim\Gamma_-\simeq\Gamma_{\rm sph}$.
Hence we have
\begin{equation}
 {\dot n}_B =  \Gamma_-\left({{\Gamma_+}\over{\Gamma_-}}-1\right)
        \simeq \Gamma_{\rm sph}({ e}^{-\mu_B/T}-1)
        \simeq -{{\Gamma_{\rm sph}\mu_B}\over T}.
   \label{eq:B-rate}
\end{equation}
\end{document}